\documentclass[a4paper,12pt]{article}
\usepackage{amssymb,amsmath,amscd,amsfonts,amsthm,mathrsfs}
\usepackage{verbatim,paralist}
\usepackage{pgflibraryarrows}
\usepackage{pgflibrarysnakes}
\usepackage{tikz}
\input xy
\xyoption{all}
\usepackage[all]{xy}
\usepackage{xcolor}
\usepackage[active]{srcltx}
\usepackage[utf8]{inputenc}
\usepackage[T1]{fontenc}
\usepackage[english]{babel}
\usepackage{ae,icomma} 

\newcommand{\DATUM}{04-11-2025}              
\newcommand{\change}
{{\marginpar{\#}}}        
\newcommand{\comma}{\: ,}              
\newcommand{\period}{\: .}             
\newcommand{\cA}{{\cal A}}

\newcommand{\cC}{{\cal C}}

\newcommand{\cT}{{\cal T}}
\newcommand{\cU}{{\cal U}}

\newcommand{\cW}{{\cal W}}

\newcommand{\field}[1]{\mathbb{#1}}
\newcommand{\R}{\field{R}}            
\newcommand{\C}{\field{C}}            

\newcommand{\rL}{{\rm L}}                 

\newcommand{\cirS}{\mathop{\bigcirc\kern -.73em {\scriptstyle{\rm S}}}}

\theoremstyle{plain}

\newcommand{\donne}{\mapsto}

\newcommand{\ic}{\mathrm{i}}

\renewcommand {\l}{\left}
\newcommand {\ri}{\right}

\newcommand{\beq}{\begin{equation}}
\newcommand{\Leq}[1]{\label{#1}\end{equation}}
\newcommand{\eeq}{\end{equation}}
\newcommand{\beqno}{\begin{eqnarray*}}
\newcommand{\eeqno}{\end{eqnarray*}}
\newcommand{\bem}{\l(\! \begin{array}}
\newcommand{\eem}{\end{array}\!\ri)}
\newcommand{\bsm}{\left(\begin{smallmatrix}} 
\newcommand{\esm}{\end{smallmatrix}\right)}  

\begin{document}

\setcounter{section}{0} 

\title{On the mathematical treatment of the Born-Oppenheimer approximation.}
\author{
{\bf Thierry Jecko}\\
AGM, UMR 8088 du CNRS, Université de Cergy-Pontoise,\\
Département de mathématiques, site de Saint Martin,\\
2 avenue Adolphe Chauvin, F-95000 Pontoise, France. \\
e-mail: thierry.jecko@u-cergy.fr, web: http://jecko.u-cergy.fr//\\}
\date{\DATUM}
\maketitle

\begin{abstract}
Motivated by the paper \cite{sw} by B.T. Sutcliffe and R.G. Woolley, we present the main ideas
used by mathematicians to show the accuracy of the Born-Oppenheimer approximation for molecules.
Based on mathematical works on this approximation for molecular bound states, in scattering theory,
in resonance theory, and for short time evolution, we give an overview of some rigorous results obtained
up to now. We also point out the main difficulties mathematicians are trying to overcome and 
speculate on further developments. The mathematical approach does not fit exactly to the
common use of the approximation in Physics and Chemistry. We criticize the latter and 
comment on the differences, contributing in this way to the discussion on the
Born-Oppenheimer approximation initiated in \cite{sw}. \\
The paper neither contains mathematical statements nor proofs. Instead we try to make accessible
mathematically rigourous results on the subject to researchers in Quantum Chemistry or Physics.

\end{abstract}






\section{Introduction.}
\label{intro}
\setcounter{equation}{0}

In the paper \cite{sw} (see also \cite{sw2,sw3}), the authors made a remarkable effort
to understand the mathematical literature on the Born-Oppenheimer approximation. It was certainly
not a easy task for them to extract relevant information for Chemistry from papers,
which often use elaborate mathematical tools and provide more or less abstract results. For
instance, they comment on the paper \cite{kmsw} that makes use of the semiclassical pseudodifferential
calculus and of an important, but rather complicated, trick (due to Hunziker in \cite{hu}) to control 
the Coulomb singularities appearing in the potential energy of the molecule. They also pointed out
to their colleagues in Chemistry some misunderstandings and too crude simplifications in the traditional
treatment of the Born-Oppenheimer approximation for molecules. One may feel a slightly pessimistic
note in the paper \cite{sw} on the possibility for Chemists to use the Born-Oppenheimer approximation
in a correct and accurate way and to benefit from mathematical works on the subject. Here we shall
give a description of the situation from the mathematical point of view, which does show the present
difficulties and limitations of the mathematical approach but still leads to a quite optimistic
impression. This is probably due to a different interpretation of the works \cite{bo} and
\cite{bh}. We shall comment in detail on this difference and also point out some questionable statements in
the traditional presentation of the approximation. \\
In the last three decades, mathematically rigorous works on the validity of the Born-Oppenheimer
approximation for molecules has been produced. We sort these works in two (incomplete) lists into
alphabetic order. The first one contains articles that, strictly speaking, study the Born-Oppenheimer
approximation: \cite{c,cds,cs,ha1,ha2,ha3,ha4,ha5,ha7,hh,hj1,hj2,hj3,hj6,hj8,hrj,jec1,jec2,jec3,
jkw,kmsw,kmw1,kmw2,ma1,ma2,ma3,ma4,mm,ms1,ms2,pst,ra,rou,st,tw}. In the second list, we mention closely
related works on semiclassical Schrödinger matrix operators: \cite{dfj,fg,fr,fln,ne,ha6,hj4,jec4,ht}.
In the first list, the papers essentially show that a reduced Hamiltonian (a Schrödinger operator
with a matrix or operator valued potential) is a good approximation to the true molecular Hamiltonian.
In the second list, the works obtain mathematical results on the reduced Hamiltonian, that are
of physical or chemical relevance for molecules. In this work, we chose to describe results that were
obtained by the use of rather involved mathematical tools and tried to make these results and the ideas
at the core of their proofs accessible for non-mathematicians. However we point out that there exist
other ways to study the Born-Oppenheimer approximation, that are easier to understand. Here we have in
particular in mind the use of coherent states (see \cite{ha1,ha2,ha3,ha4,ha5,ha6,ha7,hh,hj1,hj2,hj3,hj4,hj5,
hj6,hj7,hj8,hj9,ht,hrj}). We decided not to describe this approach in detail but we shall mention some
of its results.  \\
In the present paper, we focus on the Born-Oppenheimer approximation in the mathematical sense,
namely the possibility to approximate, for large nuclear masses, the true molecular Hamiltonian by
some effective Hamiltonian usually called the adiabatic operator. In Section~\ref{traditional-bo},
we review the original Born-Oppenheimer approximation and its actual version in Physics and Chemistry.
In Section~\ref{model}, we proceed to the removal of the centre of mass motion in two ways, one adapted
to the study of bound states and the other to scattering theory. In Section~\ref{math-view}, we
present the core of the mathematical form of the Born-Oppenheimer approximation and describe the
construction of the adiabatic Hamiltonian. In Section~\ref{comparison}, we compare this approach
with the usual interpretation and extract the main differences.
In Section~\ref{math-results}, we explain selected mathematical works
and comment on the actual difficulties and limitations of the theory. In
Section~\ref{conclusion}, we sum up the main features in the mathematical Born-Oppenheimer
approximation and argue that further progress towards chemically relevant questions
can be reasonably achieved. The last section is followed by an appendix that provides
explanations on some mathematical notions, that might be not familiar to the reader. Finally,
we added two figures used at many places in the text.\\
As pointed out in the abstract, we only present intuitive arguments and statements, that do not
respect at all the standard rigour in mathematics. But they do have a rigorous
counterpart in the mathematical literature.\\
{\bf Acknowledgement:} The author is particularly grateful to S. Golénia for the figures. He also thanks V. Georgescu, B.T. Sutcliffe,
and R.G. Woolley for fruitful discussions. The author is financially supported by the CNRS and the ANR ``NOSEVOL''.

\section{Review of the Born-Oppenheimer approximation in Physics and Chemistry.}
\label{traditional-bo}
\setcounter{equation}{0}

In this section we shortly review the framework and the meaning of the Born-Oppenheimer approximation,
as it is presented in Physics and Chemistry. Our review is based on the references \cite{bh,me,sw,w}.
We also comment on and criticize this presentation, preparing in this way the comparison with the mathematical
approach (developed in Section~\ref{math-view}) we shall perform in Section~\ref{comparison}.
For simplicity, we do not remove the motion of the centre of mass. This can however be done: see
\cite{sw}. To begin with we recall the original work \cite{bo}. Then we focus on the actual theory. 

We consider a molecule with $M$ nuclei with positive masses $m_1, m_2, \cdots , m_M$ respectively,
with positive charges $Z_1, Z_2, \cdots , Z_M$ respectively, and with $N$ electrons with mass set
equal to $1$. We set the Planck constant $\hbar$ and the electronic charge $e$ to $1$.
We denote by $(x_n)_1, \cdots , (x_n)_M$ the positions of the nuclei and the corresponding momentum 
operators by $(P_n)_1, \cdots , (P_n)_M$. We set $x_n=((x_n)_1, \cdots , (x_n)_M)$ and
$P_n=((P_n)_1, \cdots , (P_n)_M)$. Similarly the positions and momentum operators of the electrons
are denoted by $x_e=((x_e)_1, \cdots , (x_e)_N$ and $P_e=((P_e)_1, \cdots , (P_e)_N)$, respectively.
The Hamiltonian of the molecule is given by:
\begin{eqnarray}
H_{mol}&=&T_n\, +\, T_e\, +\, W\comma\label{eq:def-H-mol-physique}\\
T_n&=&\sum_{i=1}^M\frac{1}{2m_i}(P_n)_i^2\comma\hspace{.4cm}
T_e\ =\ \sum_{i=1}^N\frac{1}{2}(P_e)_i^2\comma\label{eq:kinetic-physique}\\
W&=&\sum_{\substack{i,j\in\{1,\cdots , M\}\\
i\not=j}}\frac{Z_iZ_j}{|(x_n)_i-(x_n)_j|}\, +\, \sum_{\substack{i,j\in\{1,\cdots , N\}\nonumber\\
i\not=j}}\frac{1}{|(x_e)_i-(x_e)_j|}\\
&&-\, \sum_{i=1}^{M}\sum_{j=1}^{N}\frac{Z_i}{|(x_n)_i-(x_e)_j|}\period\label{eq:interaction-physique}
\end{eqnarray}
We consider the eigenvalue problem $H_{mol}\psi =E\psi$. In \cite{bo} the authors introduce
a small parameter $\kappa$ which is an electron/nucleon mass ratio to the power $1/4$ and
consider the clamped nuclei Hamiltonian $T_e+W$, where $x_n$ is viewed as a parameter.
They assume that, for all $x_n$, there is a simple eigenvalue $\lambda (x_n)$ of the clamped
nuclei Hamiltonian for the nuclear position $x_n$. Let $x_e\donne\psi (x_n; x_e)$ be an associated
eigenvector. The set $\{\lambda (x_n); x_n\in\R^{3M}\}$ is a potential energy surface (PES).
They make the assumption of classical nature that the nuclei stay close to some isolated
``equilibrium'' position. Such a position $x_n^0$ turns out to be an isolated, local minimum of the PES.
Notice that $(x_n^0, 0)$ is also an isolated, local minimum for the nuclear, classical Hamilton function
\begin{equation}\label{eq:classical-hamiltonian}
(x_n, p_n)\ \donne\ h(x_n, p_n)=\sum_{i=1}^M\frac{1}{2m_i}(p_n)_i^2\, +\, \lambda (x_n)\period
\end{equation}
Therefore an isolated ``equilibrium'' position can be identified with an equilibrium position in
the sense of classical mechanics. Near such an ``equilibrium'' position, the authors perform
an expansion of Taylor type w.r.t. $\kappa$ of the quantum eigenvalue equation. Assuming the
existence of an expansion in power of $\kappa$ of the eigenvalue and the corresponding eigenvector,
they compute many terms of them. The eigenvalue $E$ is then close to the value of $\lambda$
at the ``equilibrium'' position. \\
Now we recall the modern version of the Born-Oppenheimer approximation in Physics and Chemistry.
We mainly follow the book \cite{bh} and add some contributions from \cite{me,sw,w}. The starting
point is again the construction of the electronic levels (or PES). One considers the clamped
nuclei Hamiltonian $T_e+W$ and regard the nuclear variables $x_n$ as parameters. Fix such a position
$x_n$. Let $(\lambda _a(x_n))_a$ be the ``eigenvalues'' of $T_e+W(x_n)$, repeated according to their
multiplicity. It is assumed that one can solve the eigenvalue problem in the sense that, for every
``eigenvalue'' $\lambda _a(x_n)$, we have an ``eigenvector'' $x_e\donne\psi _a(x_n; x_e)$.
This is done for all possible $x_n$. Based on the ``commutation'' of the clamped nuclei Hamiltonian
with the nuclear position operators and on a ``simultaneous diagonalization'' of these operators,
it is then argued that the family $(\psi _a(\cdot ; \cdot ))_a$
is ``complete'' and that the bound state $\psi$ can be recovered by a superposition of electronic
levels:
\begin{equation}\label{eq:superposition}
\psi (x_n; x_e)\ =\ \sum_a\theta _a(x_n)\psi _a(x_n; x_e)\comma
\end{equation}
It is also argued that, for each $x_n$, the family $(\psi _a(x_n ; \cdot ))_a$ can be chosen
``orthonormal'' for the scalar product in the $x_e$ variables. Inserting \eqref{eq:superposition} in
$(H_{mol}-E)\psi =0$ and taking the scalar product in the $x_e$ variables with $\psi _b$, one derives
the following coupled equations indexed by $b$:
\begin{equation}\label{eq:coupled}
(T_n\theta _b+(\lambda _b-E)\theta _b)\psi _b\, +\, \sum_ac_{a, b}\theta _a\ =\ 0\comma
\end{equation}
where the terms $c_{a, b}$ involve derivatives of the $\psi _a$ w.r.t. $x_n$ and the vector of nuclear
momentum operators $P_n$.\\
``Making the Born-Oppenheimer approximation'' in Physics corresponds to the assumption that
one may neglect the last term on the l.h.s of \eqref{eq:coupled}. The equations become uncoupled
and a bound state may be written as a product of the form $\theta _b(x_n)\psi _b(x_n; x_e)$.
A justification of such an approximation is provided in \cite{me} when the electronic levels
are simple and ``sufficiently separated''. One can also assume that a part of the contribution of the
$c_{a, b}$ is negligible (cf. the first order Born-Oppenheimer correction). \\
In \cite{w}, Equation~\eqref{eq:coupled} is derived in a slightly different way using ``states with
definite values of the electronic and nuclear coordinates''. An interesting point is a discussion of the
original result by \cite{bo} in the framework of Equations~\eqref{eq:superposition}
and~\eqref{eq:coupled}. 

In \cite{sw}, the authors expressed their dissatisfaction in the above, traditional derivation of the
Born-Oppenheimer approximation and claimed that \eqref{eq:coupled} is not properly justified. They did not
capture the main difficulties but felt one of them, namely the absence of normalized eigenvectors for
``continuous eigenvalues''. \\
If $\lambda _a(x_n)$ is a true eigenvalue, the corresponding eigenvector $x_e\donne\psi _a(x_n; x_e)$ is a square
integrable function in the $x_e$ variables that satisfies
\begin{equation}\label{eq:elec-eig}
(T_e+W(x_n))\psi _a(x_n; \cdot)=\lambda _a(x_n)\psi _a(x_n; \cdot)\period
\end{equation}
But the spectrum of $T_e+W(x_n)$ contains a continuous part and, if $\lambda _a(x_n)$ sits in this part, then Equation \eqref{eq:elec-eig}
has (most of the time) no nonzero square integrable solution. It is not explained, neither in \cite{bh} nor in \cite{me,sw,w}, in which sense
$\psi _a(x_n; \cdot )$ is an eigenvector when $\lambda _a(x_n)$ belongs to the continuous spectrum.
In \cite{w} one uses joint ``eigenvectors'' for the electronic and nuclear positions operators, which have
only continuous spectrum. There is an intuitive description of these states (see Section 3.2 in \cite{w}) and
their properties are somehow postulated, inspired by the true properties that one has when the operator has
discrete spectrum (like the harmonic oscillator). The main drawback with this notion of ``eigenvectors''
is its vagueness. One cannot check their properties, one cannot find out in which meaning any formula involving them
has to be understood. For instance, with $a$ ranging in a set having a continuous part, what is the sum in
\eqref{eq:superposition}? Even worse, the reader is not warned that there is some fuzziness at that point.
But it would be easy to write something like: `` Imagine that the element in the continuous spectrum
have eigenvectors with the following properties ...'' and then ``derive'' Equations~\eqref{eq:superposition} and~\eqref{eq:coupled}.
Or, even better, one could replace the electronic Hamiltonian by an harmonic oscillator and perform the derivation in that case.
This would at least give an idea of the Born-Oppenheimer approximation without introducing confusion. \\
It turns out that many notions and statements in the above description can be reformulated
in a clear, coherent, and rigorous way, and the main tool for that is the complicated, subtle notion of spectral
resolution (see the Appendix). In particular, the latter encodes in a correct way the influence of the continuous
spectrum. This will be seen in Section~\ref{math-view} and commented in Section~\ref{comparison}.\\
Nevertheless there is another, much deeper difficulty that might hinder the derivation of (a precise version of) \eqref{eq:coupled}.
It is the presence of ``thresholds'' in the spectrum of $T_e+W(x_n)$ between the bottom of its continuous part and
the energy $0$. In Section~\ref{comparison}, we shall try to explain the notion of threshold and its influence on \eqref{eq:coupled}.

Our latter comments seem to reinforce the pessimistic posture in \cite{sw}. But we shall see that, thanks to the mathematical
approach presented in Section~\ref{math-view}, one can avoid complicated issues concerning the continuous spectrum and
the spectral resolution and present the most useful part of the Born-Oppenheimer approximation in a rather elementary way.

\section{Removal of the mass centre.}
\label{model}
\setcounter{equation}{0}

In this section, we prepare the presentation of the mathematical version of the
Born-Oppenheimer in Section~\ref{math-view}. We recall the framework and perform the
removal of the centre of mass.

For convenience, we change the notation of the variables and 
rewrite \eqref{eq:def-H-mol-physique}, \eqref{eq:kinetic-physique}, and
\eqref{eq:interaction-physique}. Denoting by $z_1, z_2, \cdots , z_M\in\R^3$ the positions of
the nuclei and by $z_{M+1}, z_{M+2}, \cdots , z_{M+N}\in\R^3$ the positions of the electrons, the
Hamiltonian of the molecule is given by:
\begin{eqnarray}
H_{mol}&=&K\, +\, W\comma\label{eq:def-H-mol}\\
K&=&-\sum_{i=1}^M\frac{1}{2m_i}\Delta_{z_i}-
\frac{1}{2}\sum_{j=M+1}^{M+N}\Delta_{z_j}\comma\label{eq:kinetic}\\
W&=&\sum_{\substack{i,j\in\{1,\cdots , M\}\\
i\not=j}}\frac{Z_iZ_j}{|z_i-z_j|}\, +\, \sum_{\substack{i,j\in\{M+1,\cdots , M+N\}\nonumber\\
i\not=j}}\frac{1}{|z_i-z_j|}\\
&&-\, \sum_{i=1}^{M}\sum_{j=M+1}^{M+N}\frac{Z_i}{|z_i-z_j|}\period\label{eq:interaction}
\end{eqnarray}
Here $-\Delta_{z_k}$ denotes the Laplace operator in the $z_k=(z_k^1, z_k^2, z_k^3)$ variable, that is
\[-\Delta_{z_k}\ =\ -\nabla_{z_k}^2\ =\ \bigl(-\ic \partial _{z_k^1}\bigr)^2\, +\, \bigl(-\ic \partial _{z_k^2}\bigr)^2
\, +\, \bigl(-\ic \partial _{z_k^3}\bigr)^2\comma\]
where $\partial _t$ stands for the partial derivative with respect to the variable $t$. 

It is usual and physically relevant to remove from the Hamiltonian $H_{mol}$ the motion of the
centre of mass of the molecule. This is done by an appropriate change of variables. There is no
canonical choice for this change of variables. This means in particular that one can choose it
according to the kind of study one wants to perform. To study bound states of the molecule or
its time evolution, we shall use the change of variables adopted in \cite{kmsw,ms2,sw}.
To consider diatomic collisions (ion-ion, ion-atom, or atom-atom scattering), we shall use
another one (those in \cite{kmw1,kmw2}); see p. 75-82 in \cite{rs3}
for details on the removal of the centre of mass.

In the first mentioned situation, we take the nuclear centre of mass (which is close to the centre of
mass of the molecule), Jacobi coordinates for the nuclei, and atomic coordinates for the electrons.
Let $\cC :(z_1, \cdots , z_{M+N})\mapsto (R; x_1, \cdots , x_{M-1}, y_1, \cdots , y_N)$ be the change
of variables defined by 
\begin{eqnarray*}
&&m\ =\ \sum _{k=1}^Mm_k\comma\hspace{.4cm}R\ =\ \frac{1}{m}\sum _{k=1}^Mm_kz_k\comma\hspace{.4cm}
\mbox{for}\, 1\leq j\leq N\comma\ y_j\ =\ z_{M+j}-R\comma\\
&&\mbox{for}\, 1\leq j\leq M-1\comma\ x_j=z_{j+1}-\frac{1}{\sum _{k\leq j}m_k}\sum_{k\leq j}m_kz_k
\period
\end{eqnarray*}
$R$ is the centre of mass of the nuclei, the $x_j$ are the new ``nuclear'' coordinates, and the
$y_j$ are the new electronic variables. For an appropriate constant $C$ that only depends on
the masses and on $N$ we define, for any $\rL^2$ function $f$ of the variables $(z_1,
\cdots , z_{M+N})$, 
\begin{equation}\label{eq:change-variables}
 (\cU f)(R; x_1, \cdots , x_{M-1}, y_1, \cdots , y_N)\ =\ Cf\bigl(\cC ^{-1}(R; x_1, \cdots , x_{M-1},
y_1, \cdots , y_N)\bigr)\period
\end{equation}
The constant $C$ is chosen such that, for all $f$, $f$ and $\cU f$ have the same $\rL^2$-norm
($\cU$ is unitary), keeping unchanged the physical interpretation of the $\rL^2$-norm.
Looking at the Hamiltonian $H_{mol}$ in the new variables $(R; x_1, \cdots x_{M-1}, y_1, \cdots ,
y_N)$ means that we consider the operator
\begin{equation}\label{eq:change-unitaire}
\cU H_{mol} \cU ^{-1}\ =\ -\frac{1}{2m}\Delta _R\, +\, H\comma
\end{equation}
where $H$ only acts on the variables $(x_1, \cdots , x_{M-1}, y_1, \cdots , y_N)$. Forgetting
about the kinetic energy operator of the nuclear centre of mass, we focus on the physically
relevant Hamiltonian $H$. Denoting by $\nabla _t$ the gradient operator in the variable $t$ and
setting, 
\begin{equation*}
\mbox{for}\, 1\leq j\leq M-1\comma\ \mu_j^{-1}\, =\, m_{j+1}^{-1}+\Bigl(\sum _{k\leq j}m_k\Bigr)^{-1}
\comma\ x=(x_1, \cdots , x_{M-1})\comma\ \mbox{and}\ y=(y_1, \cdots , y_{N})\comma
\end{equation*}
the latter is given by
\begin{eqnarray}
H&=&H_0\, +\, T_{\rm HE}\comma
\label{eq:def-H}\\
H_0&=&-\sum _{j=1}^{M-1}\frac{1}{2\mu _j}\Delta _{x_j}\, +\, Q\comma
\label{eq:def-H_0}\\
T_{\rm HE}&=&\sum_{1\leq k<j\leq N}c_{kj}\nabla _{y_k}\cdot\nabla _{y_j}\comma
\label{eq:def-hughes-eckart}\\
Q&=&\int^{\oplus}Q(x)\, dx\comma\hspace{.2cm}Q(x)\ =\ -\sum _{k=1}^N\Delta _{y_k}\, +\, W(x; y)
\label{eq:def-Q}
\end{eqnarray}
where $W(x; y)$ is just the function $W$ in \eqref{eq:interaction} composed with the inverse change
of variables $\cC^{-1}$. The action of the direct integral operator $Q$
on a function $\psi (x; y)$ gives another function $\tau (x; y)$ defined, for all $x$, by
$\tau (x; y)=Q(x)\psi (x; y)$, that is the action of the operator $Q(x)$ on the function
$y\donne \psi (x; y)$. Details on the notion of direct integral are provided in the Appendix. \\
Compared to Section~\ref{traditional-bo}, the $x$ (resp. $y$) variables play the
rôle of the $x_n$ (resp. $x_e$) variables. We observe that $Q(x)$ is an operator in the electronic $y$
variables that depends only parametrically on the nuclear $x$ variables and does not depend on $R$.
The operator $T_{\rm HE}$ is usually called the Hughes-Eckart term. For each nuclear configuration
$x$, the operator $Q(x)$ is referred to as the electronic Hamiltonian in the configuration $x$ (it is
called the clamped-nuclei Hamiltonian in \cite{sw}).
Note that the coefficients $\mu _j$ in $H$ are missing in \cite{kmsw} (in fact they were moved into the interaction terms). This has no consequence on the validity of the results in this paper, that also hold true for the present Hamiltonian $H$.

Next we turn to the scattering situation. For simplicity, we restrict ourselves to the
diatomic case (i.e. $M=2$). We still look at the Hamiltonian $H_{mol}$ but we want now to describe
the collision of two ions (or two atoms, or an atom and an ion). It is useful to choose a change
of variables that allows an easy description of the system at the beginning of the collision process
(and another one, to describe the system after the collision). To this end, we introduce a cluster
decomposition $c=\{c_1', c_2'\}$ with  $c_j'=\{j\}\cup c_j$, for $j=1,2$, and $c_1$, $c_2$ form a
partition of the set $\{3,\cdots , N+2\}$. At the beginning of the scattering process, the particles
are gathered in two clusters described by $c_1'$ and $c_2'$. Each cluster contains a nucleus
illustrating the fact that we consider a collision of two ions (and not a collision of some electrons
with a molecule). Since the motion of the centre of the system is not relevant for scattering, we
shall remove it. In order to do so, we use the particular change of variables in \cite{kmw1,kmw2},
which also allows a good description of the scattering processes associated to the decomposition $c$. \\
For $k\in\{1; 2\}$, denote by $|c_k|$ the number of electron in the cluster $k$. Its mass is then
$M_k:=m_k+|c_k|$ and its mass centre is located at:
\begin{equation}\label{mass-centre-cluster}
 R_k:=\frac{1}{M_k}\Bigl(m_kz_k+\sum _{j\in c_k}z_j\Bigr)\period
\end{equation}
In particular, the total mass $m$ of the molecule is $M=m_1+m_2+N=M_1+M_2$.
The new variables are, for $k\in\{1; 2\}$,
\begin{equation}\label{ch-var-kmw}
R:=\frac{1}{M}\Bigl(m_1z_1+m_2z_2+\sum _{j=3}^{N+2}z_j\Bigr)\, , \,
x:=R_1-R_2\, ,\, y _j:=z_j-z_k\, , \mbox{for}\ j\in c_k\period
\end{equation}
We set, for $y\in\R^{3N}$,
\begin{equation}\label{linear-map-l}
\ell (y):=\frac{1}{M_1}\sum _{j\in c_1}y_j-\frac{1}{M_2}\sum _{j\in c_2}y_j\period
\end{equation}
As above, this change of variables $\cC _c$ induces a unitary map $\cU _c$ by
\begin{equation}\label{eq:change-variables-c}
 (\cU _cf)(R; x, y_1, \cdots , y_N)\ =\ C_cf\bigl(\cC _c^{-1}(R; x, y_1, \cdots , y_N)\bigr)\period
\end{equation}
Looking at $H_{mol}$ in the new variables amounts to considering
\begin{equation*}
\cU _cH_{mol} \cU _c^{-1}\ =\ -\frac{1}{2m}\Delta _R\, +\, H'\comma
\end{equation*}
where $H'$ only acts on the variables $(x, y_1, \cdots , y_N)$. Taking away the motion of the
centre of mass of the full system again, we keep our attention on $H'$, which is given by
\begin{eqnarray}
H'&:=&-\frac{1}{2}\Bigl(\frac{1}{M_1}+\frac{1}{M_2}\Bigr)\Delta _x\, +\, T_{\rm HE}'\, +\, Q_c
\comma\label{eq:H'}\\
T_{\rm HE}'&=&-\sum_{k=1}^2\frac{1}{2m_k}\Bigl|\sum _{j\in c_k}\ic \nabla_{y_j}\Bigr|^2\\
Q_c&=&\int^{\oplus}Q_c(x)\, dx\comma\hspace{.4cm}I_c\ =\ \int^{\oplus}I_c(x)\, dx\comma
\label{direct-int-c}\\
Q _c(x)&:=&Q^c+I_c(x)\comma\label{eq:Q_c}\\
Q ^c&:=&\sum_{k=1}^2\left\{\sum_{j\in c_k}\Bigl(-\frac{1}{2}\Delta_{y_j}-
\frac{Z_k}{|y_j|}\Bigr)+\sum_{i,j\in c_k, i\not=j}\frac{1}{|y_i-y_j|}\right\}\comma\label{eq:Q^c}\\
I_c(x)&:=&-\sum_{j\in c_1}\frac{Z_2}{|y_j+x-\ell(y)|}-\sum_{j\in c_2}\frac{Z_1}{|y_j-x+\ell(y)|}
+\frac{Z_1Z_2}{|x-\ell(y)|}\label{eq:Ic-kmw}\\
&+&\sum_{i\in c_1,\, j\in c_2}\frac{1}{|y_i-y_j+x-\ell(y)|}+\sum_{i\in c_2,\, j\in c_1}
\frac{1}{|y_i-y_j-x+\ell(y)|}\comma\nonumber\\
H_c'&:=&H'\, -\, I_c\period\label{eq:H_c'}
\end{eqnarray}
$Q^c$ stands for the Hamiltonian of separated (noninteracting) clusters while $I_c$ contains all
extracluster interactions. The electronic Hamiltonian for the nuclear position $x$ is 
$Q_c(x)$. The term $T_{\rm HE}'$ is the Hughes-Eckart term in this situation. As usual in scattering
theory, we should mention the reference (or free) dynamics to which the full dynamics has to be
compared to (for time tending to $-\infty$). The free dynamics is generated by the Hamiltonian $H_c'$
in \eqref{eq:H_c'}, that is, up to the Hughes-Eckart term $T_{\rm HE}'$, the Hamiltonian of freely
moving clusters $c_1$ and $c_2$. If the scattering process under consideration produces at the
end (i.e. for time tending to $+\infty$) a cluster decomposition $d$ of the system then the
free dynamics for large positive times is given by $H_d'$.

Using Kato's perturbation argument and Hardy's inequality (cf. \cite{k,rs2}), one can show that
$H$ can be realized as a self-adjoint operator. This means that $H$ can be defined on an appropriate
subspace of the $\rL^2$ functions of the variables $(x_1, \cdots , x_{M-1}, y_1, \cdots ,
y_N)$ and the resulting operator is self-adjoint. In fact, this subspace is the domain of
the Laplace operator in all variables $(x_1, \cdots , x_{M-1}, y_1, \cdots , y_N)$. The same holds 
true for $H_0$. In the same way,
$H_{mol}$ (respectively $H'$) is self-adjoint on the domain of the Laplace operator of
all variables $z_{1}, \cdots , z_{M+N}$ (respectively $x, y_1, \cdots , y_N$). For fixed $x$,
$Q(x)$ (respectively $Q_c(x)$), as an operator on the variables $y_1, \cdots , y_N$, also has
a self-adjoint realization by the same argument. Since $Q$ (respectively $Q_c$), as an operator
on the variables $x, y_1, \cdots , y_N$, is a direct integral of self-adjoint
operators, it has also a self-adjoint realization (cf. p. 284 in \cite{rs4}).

\section{Mathematical approach to the Born-Oppenheimer approximation.}
\label{math-view}
\setcounter{equation}{0}

Here we want to present the main ideas behind the mathematical treatment of the
Born-Oppenheimer approximation, which was initiated in \cite{c,cds,cs}. Since its validity
should rest on the fact that
the nuclei are much heavier than the electrons, one introduces a small, positive parameter $\epsilon$
related to the electron/nucleon mass ratios. For instance, in \cite{kmsw}, the nuclear masses
$m_k$ are given by $m_k=\epsilon^2\lambda _k$, where the $\lambda_k$ are of order $1$, and, in
\cite{kmw1,kmw2}, one uses $\epsilon^2=M_1^{-1}+M_2^{-1}$ (with the notation of Section~\ref{model}).
Anyhow the main point is that $\epsilon$ is always sent to $0$. This means that the results proved
hold true for ``small enough'' $\epsilon$ and, most of the time, one has no concrete idea of how
small $\epsilon$ should be. This restriction is of course a drawback for physical or chemical purposes
but it is useful to understand the small $\epsilon$ limit and it often gives correct results when
compared with the observed behaviour of the physical system. \\
Now we come to the main features that ensure the validity of the Born-Oppenheimer approximation.
Let us consider a normalized, bound state $\varphi$ of energy $E$ of the operator $H$ in
\eqref{eq:def-H}. We note that
$\mu _j^{-1}=\epsilon^2\mu _j'^{-1}$ for $\epsilon$-independent $\mu _j'$. Let us temporarily fix the
nuclear variable $x$. Typically the spectrum of $Q(x)$ starts with some isolated
eigenvalues $\lambda _1(x), \cdots , \lambda _J(x)$ with finite multiplicity $k_1, \cdots , k_J$
respectively and has above a
continuous part $\sigma _c(Q(x))$ (see fig. 1 in the diatomic case). Here $J$ could be infinite, i.e.
one may have an infinite number of eigenvalues below $\sigma _c(Q(x))$. Since $Q(x)$ is self-adjoint, 
we can decompose $\varphi (x; \cdot)$ in a ``basis'' of electronic ``eigenvectors'' of $Q(x)$:
\begin{equation}\label{eq:decomp-phi}
\varphi (x, \cdot )\ =\ \sum_{j=1}^J\sum _{k=1}^{k_j}\langle \psi _{jk}(x; \cdot ), \varphi (x; \cdot)\rangle _y
\psi _{jk}(x; \cdot )\, +\, \int _{\lambda\geq\inf\sigma _c(Q(x))}dS_x(\lambda )\varphi (x; \cdot)\,
\comma
\end{equation}
where the $\psi _{jk}(x; \cdot )$ form a basis of true eigenvectors of $Q(x)$ associated to
$\lambda _j(x)$ respectively, $\langle \cdot, \cdot\rangle_y$ denotes the usual scalar product in the $y$ variables, and
$dS_x(\lambda )$ is the (restriction to the continuous spectrum of the) spectral resolution of $Q(x)$. If $J$ is infinite, the first term in \eqref{eq:decomp-phi} has to be understood, in an appropriate way, as the sum of a series.  See ``Spectral resolution'' in the Appendix for details. \\
All this can be done for all values of the variable $x$.\\
We assume that $E$ belongs to a small open energy interval $(E_-; E_+)$ below the infimum over
all $x$ of $\sigma _c(Q(x))$ (as in fig. 1). Since the coefficients $c_{kj}$ in
\eqref{eq:def-hughes-eckart} contain $\epsilon^2$, the Hughes-Eckart term $T_{\rm HE}$ is small compared
to the electronic Hamiltonian $Q$ in \eqref{eq:def-Q}. To compute $E$ and $\varphi$, the idea is
that only the part of the spectrum of the operators $Q(x)$ less or equal to $E_+$ should be relevant,
since the nuclear kinetic energy is nonnegative. Define
$J_+$ as the largest $j\leq J$ such that there exists some $x$ with $\lambda _j(x)\leq E_+$. It turns out that 
$J_+$ is finite. In \eqref{eq:decomp-phi}, we expect that
\begin{equation}\label{eq:approx}
\varphi (x, \cdot )\ =\ \sum_{j=1}^{J_+}\sum _{k=1}^{k_j}\langle \psi _{jk}(x; \cdot ),
\varphi (x; \cdot)\rangle _y\psi _{jk}(x; \cdot )\, +\, \mbox{small term}\period
\end{equation}
Let $\Pi (x)$ be the orthogonal projection on the
first $J_+$ energy levels of $Q(x)$ that is, for an electronic wavefunction $\psi (y)$,
\begin{equation}
\Pi (x)\psi\ =\ \sum_{j=1}^{J_+}\sum _{k=1}^{k_j}\langle \psi _{jk}(x; \cdot ), \psi\rangle _y
\psi _{jk}(x; \cdot )\period\label{eq:def-pi}
\end{equation}
We set $\Pi ^\perp (x)=1-\Pi (x)$. Note that \eqref{eq:decomp-phi} may be rewritten as
\begin{equation}\label{eq:phi-Pi}
\varphi (x, \cdot )\ =\ \Pi (x)\varphi (x, \cdot )\, +\, \Pi ^\perp (x)\varphi (x, \cdot )\comma
\end{equation}
where 
\begin{equation*}
\Pi ^\perp (x)\varphi (x, \cdot )\ =\ \sum_{j=J_++1}^J\sum _{k=1}^{k_j}\langle \psi _{jk}(x; \cdot ), \varphi (x; \cdot)\rangle _y
\psi _{jk}(x; \cdot )\, +\, \int _{\lambda\geq\inf\sigma _c(Q(x))}dS_x(\lambda )\varphi (x; \cdot)\period
\end{equation*}
To implement this idea, it is natural to try as an effective, self-adjoint Hamiltonian an operator
of the form $\Pi G\Pi$ where $G$ may be $H$ or $H_0$ and $\Pi$ is the direct integral of the $\Pi (x)$.
It acts on a total wavefunction $\psi (x, y)$
as follows: for each $x$, one projects $\psi (x, \cdot)$ as in
\eqref{eq:def-pi} then one lets $G$ act on the result and, finally, one projects again according
to \eqref{eq:def-pi}. Thus, the spectral subspaces of the operators $Q(x)$ corresponding to energies
above $E_+$ are removed. Such an operator is usually called an adiabatic operator.
We expect that among the eigenvalues in $(E_-; E_+)$ and corresponding
eigenvectors of $\Pi G\Pi$, there is a good approximation of $E$ and $\varphi$.
Let us try to justify intuitively this claim.\\
Assume that $(E_-; E_+)$ is a small interval around the infimum of the function $\lambda_1$ that
is attained in some region $\Gamma$ (in the diatomic case, $\lambda _1$ only depends on the norm of
$x$ and $\Gamma$ can be a sphere) and that $J_+=1$ and $k_1=1$ (for instance $(E_-; E_+)=(E_-^0; E_+^0)$ in
fig. 1). Set $\psi _1:=\psi _{11}$ for simplicity. Using the positivity of the nuclear kinetic energy,
$H\geq Q+T_{\rm HE}$ thus, by the properties of direct integrals (see the Appendix) and the identity $\Pi (x)+\Pi (x)^\perp =1$,
\begin{eqnarray*}
E=\langle \varphi , H\varphi\rangle&\geq&\int \langle \varphi (x, \cdot ), Q(x)\Pi (x)
\varphi (x, \cdot )\rangle_y\, dx\\
&&+\, \int \langle \varphi (x, \cdot ), Q(x)\Pi ^\perp(x)\varphi (x, \cdot )\rangle_y\, dx\\
&&+\, \int \langle \varphi (x, \cdot ), T_{\rm HE}\varphi (x, \cdot )\rangle_y\, dx\period\\
\end{eqnarray*}
Since $T_{\rm HE}$ is small compared to $Q$, one can show via Kato's perturbation theory that
\begin{equation}\label{eq:size-hughes-eckart}
\int \langle \varphi (x, \cdot ), T_{\rm HE}\varphi (x, \cdot )\rangle_y\, dx\ =\ O(\epsilon^2)
\ \mbox{and}\ \bigl\|T_{\rm HE}\varphi\bigr\|\ =\ O(\epsilon^2)\comma
\end{equation}
where $\|\cdot\|$ is the $\rL^2$-norm in the variables $(x, y_1, \cdots , y_N)$. 
Therefore, since $\lambda _1(x)$ is simple, 
\begin{eqnarray}
E&\geq& \int \lambda _1(x)\bigl|\langle \varphi (x, \cdot ), \psi _{1}(x; \cdot )\rangle _y\bigr|^2\,
dx\nonumber\\
&&+\, E_+\int \|\Pi ^\perp(x)\varphi (x, \cdot )\|_y^2\, dx\, +\, O(\epsilon^2)\comma
\label{eq:int}
\end{eqnarray}
where $\|\cdot\|_y$ is the $\rL^2$-norm in the $y$ variables. Here we used the fact that,
on the range of $\Pi ^\perp(x)$, $Q(x)\geq E_+$. Since $E\leq E_+$ and, by \eqref{eq:phi-Pi} and the
properties of direct integrals, 
\begin{equation}\label{eq:norm}
1\ =\ \|\varphi \|^2\ =\ \int\bigl|\langle \psi _1(x; \cdot ), \varphi (x, \cdot )\rangle _y
\bigr|^2\, dx\, +\, \int\|\Pi ^\perp(x)\varphi (x, \cdot )\|_y^2\, dx\comma
\end{equation}
the second integral in \eqref{eq:int} and the term $\bigl|\langle\psi _1(x; \cdot ), \varphi (x, \cdot )
\rangle _y\bigr|^2$ for $x$ far from $\Gamma$ should be small. In particular, in \eqref{eq:decomp-phi},
we should have
\begin{equation*}
\varphi (x, \cdot )\ =\ \chi (x)\langle \psi _1(x; \cdot ), \varphi (x; \cdot)\rangle _y
\psi _1(x; \cdot )\, +\, \mbox{small term}\comma
\end{equation*}
where $\chi$ is the characteristic function of a neighbourhood of $\Gamma$. The relevant part of
$\varphi$ is then its projection onto the electronic level $\psi _1(x; \cdot )$. Since the
lower bound in \eqref{eq:int} is close to $E$ and does not contain the nuclear kinetic energy,
the latter must be small.\\
Note that this is not sufficient if the interval $(E_-; E_+)$ is placed much higher (as in fig. 1).
Let us now produce a better argument in this more general situation. Using the second estimate in
\eqref{eq:size-hughes-eckart}, one can show that, close to $E$ and $\varphi$ respectively,
there are an energy $E_0$ and a normalized $\rL^2$-function $\varphi _0$ such that $H_0\varphi _0=
E_0\varphi _0$. We assume $E_0\in (E_-; E_+)$. In particular, $\Pi (H_0-E_0)\varphi _0=0$ and
$\Pi ^\perp (H_0-E_0)\varphi _0=0$. Using $\Pi+\Pi^\perp =1$ and $\Pi^2=\Pi$,
\begin{eqnarray}
(\Pi H_0\Pi -E_0)\Pi\varphi _0&=&-\Pi H_0\Pi ^\perp\varphi _0\comma
\label{eq:pi}\\
(\Pi ^\perp H_0\Pi ^\perp -E_0)\Pi ^\perp\varphi _0&=&-\Pi ^\perp H_0\Pi\varphi _0\period
\label{eq:pi-perp}
\end{eqnarray}
Since $\Pi (x)^\perp\Pi (x)=0$ and $\Pi (x)$ commutes with $Q(x)$,
\begin{equation*}
\Pi H_0\Pi^\perp\ =\ -\Pi \Bigl(\sum _{j=1}^{M-1}\frac{1}{2\mu _j}\Delta _{x_j}\Bigr)\Pi^\perp \ =\
\epsilon^2\sum _{j=1}^{M-1}\frac{1}{2\mu _j'}\Pi [\Delta _{x_j}, \Pi]\period
\end{equation*}
Now $\epsilon^2$ times the commutator $[\Delta _{x_j}, \Pi (x)]$ equals
$2\epsilon(\nabla_{x_j}\Pi)(x)\cdot
\epsilon\nabla _{x_j}+\epsilon^2(\Delta _{x_j}\Pi )(x)$. Since
\begin{equation*}
\int \frac{1}{2\mu _j}\bigl|\nabla _{x_j}\varphi _0\bigr|^2\, dx\ \leq \ \int \sum_{k=1}^{M-1}
\frac{1}{2\mu _k}\bigl|\nabla _{x_k}\varphi _0\bigr|^2\, dx
\end{equation*}
and the nuclear kinetic energy remains bounded (this is due to the self-adjointness of $H_0$ on
the domain of the Laplace operator and to the finiteness of the total energy), the right hand sides of \eqref{eq:pi} and \eqref{eq:pi-perp}
are $O(\epsilon)$ in $\rL^2$-norm. Since $\Pi ^\perp H_0\Pi ^\perp\geq E_+$ on the range of
$\Pi ^\perp$, the operator $\Pi ^\perp H_0\Pi ^\perp -E_0$ restricted to this range is invertible
with bounded inverse and $\eqref{eq:pi-perp}$ shows that $\Pi ^\perp\varphi _0$ is $O(\epsilon)$
in $\rL^2$-norm.
In particular, $\Pi\varphi _0$ is almost an eigenfunction of $\Pi H_0\Pi$ with $\rL^2$-norm close
to $1$ (cf. \eqref{eq:norm}) so is close to a true normalized eigenfunction of $\Pi H_0\Pi$. Thus $\varphi _0$ (and also
$\varphi$) should be also close to a normalized eigenfunction of $\Pi H_0\Pi$. Notice that, by
\eqref{eq:pi}, the nuclear kinetic energy in the state $\Pi\varphi _0$ is close to 
\begin{equation*}
-\int \langle \Pi\varphi _0\, ,\, (Q(x)-E_0)\Pi\varphi _0\rangle _y\, dx
\end{equation*}
which has no reason to be small in general. Indeed, if $E_0$ is clearly above the infimum of
$\lambda _1$ and below the infimum of $\lambda _2$, and $J_+=1$, this term equals
\begin{equation*}
-\int (\lambda _1(x)-E_0)\bigl|\langle \psi _1(x; \cdot ), \varphi _0(x, \cdot )\rangle _y\bigr|^2\, dx
\end{equation*}
and it is possible to show that the squared function essentially lives in the bottom of the ``well'':
$\{x;\lambda _1(x)\leq E_0\}$. \\
In the above argument, we used the fact that we can differentiate $\Pi (x)$ twice with respect to the variable
$x$. This is not obvious at all when one looks at the $x$-dependence in $Q(x)$ (see \eqref{eq:def-Q})
which involves the rather irregular function $W$ in \eqref{eq:interaction}. Thanks to a trick
due to Hunziker in \cite{hu}, one can prove that $x\mapsto\Pi (x)$ is smooth away from the set of
the nuclear collisions (this is actually sufficient for our argument above since one can show, by
energy arguments, using the repulsive nature of nuclear interaction, that the wavefunction is
concentrated away from these collisions). This trick is also used in \cite{kmsw} and is partially
responsible for the technical complications there.
The idea is to perform a $x$-dependent change of variables on the $y$ variables in $Q(x)$
that makes the $x$-dependent singularities in the function $W(x, y)$ in \eqref{eq:def-Q}
$x$-independent. This does not change the $\Pi (x)$.
This can be done only locally in $x$ i.e. for $x$ close enough to any fixed position
$x_0$ (see Lemma 2.1 in \cite{jec5} for details). It is essential that the $\lambda _j(x)$ are separated
from the rest of the spectrum of $Q(x)$ (in particular from its continuous part). This implies the
finiteness of the rank of $\Pi (x)$ but this property is not used. \\
The regularity of $\Pi (x)$ allows the construction of globally defined, smooth functions
$x\mapsto\psi _{jk}(x, \cdot)$ (for $1\leq j\leq J_+$ and $1\leq k\leq k_j$) with values in the
$\rL^2$ electronic functions such that, for each $x$, the family 
$(\psi _{jk}(x, \cdot))_{jk}$ is a orthogonal basis of the range of the
projection $\Pi (x)$. Because of the possible presence of eigenvalue crossing (see fig. 1),
it is not always possible to choose these functions $\psi _{jk}(x, \cdot)$ among the eigenvectors
of $Q(x)$. \\
We have seen that, up to an error of size $O(\epsilon)$, one can reduce the eigenvalue problem for $H$
to the one for $\Pi H_0\Pi$. One can compute explicitly $\Pi H_0\Pi$ in terms of the electronic
wavefunctions $\psi _{jk}(x, \cdot)$ (see \cite{pst}) and again
remove terms that are also $O(\epsilon)$. Writing the approximate eigenstate $\tilde\varphi$ of
energy $\tilde E$ as 
\begin{equation}\label{eq:born-huang-form}
\tilde\varphi\ =\ \sum_{j=1}^{J_+}\sum _{k=1}^{k_j}\theta _{jk}(x)\psi _{jk}(x; \cdot )\comma
\end{equation}
and letting act $\Pi H_0\Pi-\tilde{E}$, one ends up, in the diatomic case for simplicity, with the uncoupled equations $-\epsilon^2\Delta _x
\theta _{jk}+\lambda_j\theta_{jk}=\tilde E_{jk}\theta_{jk}$, $\tilde E_{jk}$ close to $\tilde E$, if
the $\psi _{jk}(x; \cdot )$ are eigenvectors of $Q(x)$. If the latter is not true (it can be the 
case when crossing eigenvalues occur), one has coupled partial differential equations in the $x$
variables for the $\theta _{jk}$ with coefficients depending on the
$\langle\psi _{jk}, Q(x)\psi _{j'k'}\rangle_y$. These differential equations play the rôle of
\eqref{eq:coupled} in the physical setting.\\
Now, if we demand an accuracy of $O(\epsilon^2)$, then $\Pi H_0\Pi$ still provides a good
approximation if the rank of $\Pi (x)$ is one. Taking the electronic wave function $\psi _1$ real, one 
cancels the term containing $\Pi (x)(\nabla_{x_j}\Pi)(x)$ in $\Pi H_0\Pi^\perp$, producing an error of order
$\epsilon^2$. But now, less terms in $\Pi H_0\Pi$ can be removed.
In particular, one has to keep terms containing the so-called Berry connection, i.e. factors of
the form $\langle\psi _{jk}, \nabla _{x_p}\psi _{j'k'}\rangle_y$. We see that the variation of the
$\psi _{j'k'}$ (or of $\Pi (x)$) has to be taken into account. We refer to \cite{pst} for details. \\
If one wants to improve the accuracy to $O(\epsilon^3)$ (or better), one needs to include
the Hughes-Eckart term $T_{\rm HE}$ in $Q(x)$ and to replace $\Pi (x)$ by an appropriate refined projector
$\Pi _r(x)$ which is essentially of the form $\Pi (x)+\epsilon\Pi _1(x)+\epsilon^2\Pi _2(x)$. Then one uses
as adiabatic operator $\Pi _rH\Pi _r$ (cf. \cite{ms2,pst}). It is important to note that $\Pi _r$ does depend on electronic states the energy of which are above $E_+$. \\
In the scattering situation mentioned in Section~\ref{model}, one can choose the total energy $E$
in an energy range $(E_+^c, E_-^c)$ like in fig. 2 and take $\Pi H'\Pi$ as an
effective Hamiltonian. In this case, it is important to choose the projections
$\Pi (x)$ as spectral projections of the operators $Q_c(x)+T_{\rm HE}'$ (because the Hughes-Eckart
term $T_{\rm HE}'$ has no decay in $x$ at infinity).

\section{Comparison.}
\label{comparison}
\setcounter{equation}{0}

Having presented the physical and mathematical versions of the Born-Oppenheimer approximation
in Sections~\ref{traditional-bo} and~\ref{math-view} respectively, we are now ready to compare
them. After a remark on a minor difference, we shall explain how to partially cure the vagueness of
the traditional approach and then focus on the comparison.

While, in Physics, the small parameter is essentially an electron/nucleon mass ratio to the
power $1/4$, we chose here more or less the square root of this ratio. This comes from the fact that
mathematicians see the Born-Oppenheimer approximation as a semiclassical analysis w.r.t. the nuclear
variables and, in this framework, it is convenient to consider ``$\epsilon$-derivatives'':
$\epsilon\nabla _x$ in the nuclear variables. Since we fixed the total energy, we consider, roughly speaking,
$-\epsilon^2\Delta_x$ of order one. So is also $\epsilon\nabla _x$. In particular, a term like
$2\epsilon(\nabla_{x_j}\Pi)(x)\cdot\epsilon\nabla _{x_j}$ is of order $\epsilon$ (and this
can be shown rigorously, in an appropriate sense). This gives a way to detect negligible
terms. Of course, the choice of $\epsilon$ is a matter of taste, since
we could have replaced everywhere $\epsilon$ by $\epsilon^2$.\\

First of all, we have to make an important correction in the physical presentation in Section~\ref{traditional-bo}.
One can indeed show that the operator $H_{mol}$ has no eigenvalue! (See the Appendix for a precise definition of
eigenvalues.) Let us give here an intuitive explanation of this fact. From \eqref{eq:change-unitaire}, we see that
$H_{mol}$ describes a freely moving system of particles, the internal structure of which is governed by the Hamiltonian
$H$. In such situation, one physically expects no bound states for the full system. The internal system however may have
some and this can be proved. In Section~\ref{math-view}, we actually studied this internal system. Now our presentation
in Section~\ref{traditional-bo}, which recalls the traditional one, is not completely wrong. In fact, we can say that
$H_{mol}$ has an ``eigenvalue'' $E$ if there is a nonzero solution $\psi$ of $H_{mol}\psi=E\psi$ such that the restriction
of $\psi$ to the hyperplane
\[\sum_{i=1}^M m_i(x_n)_i\, +\, \sum_{i=1}^N(x_e)_i\ =\ 0\]
has a finite $\rL^2$-norm. Notice that this hyperplane is formed by all configurations of the particles such that
the total mass centre sits at $0$. With this correction, the description of the Born-Oppenheimer approximation
in Section~\ref{traditional-bo} is meaningful. \\
The presentations in Sections~\ref{traditional-bo} and~\ref{math-view} are slightly
different since we removed the centre of mass motion in the latter and not in the former. However
this will not influence the discussion below. For their comparison, we shall use the framework of
Section~\ref{math-view} making the following identifications: $x_n\leftrightarrow x$, $x_e\leftrightarrow y$,
$T_e+W\leftrightarrow Q$, $T_e+W(x_n)\leftrightarrow Q(x)$, $T_n\leftrightarrow H_0-Q$. \\
In \cite{bh,w}, a ``complete, orthogonal set of eigenfunctions'' is considered and used to
express the total wave function $\varphi$ as a superposition of such ``eigenfunctions''.
In this set, there are true eigenfunctions and their orthogonality is expressed in term of the
$\rL^2$ scalar product. As in the well-known case of the hydrogen atom, the set of true eigenfunctions
is not complete. One has to add a contribution of the continuous spectrum.
But it is not clear in \cite{bh,w} what are the (generalized) eigenfunctions associated
to energies in the continuous spectrum. The properties of this set (completeness and orthogonality)
are deduced from the self-adjointness of the electronic Hamiltonian, viewed in the full
$\rL^2$-space, and its commutation with the nuclear position operators. Note that the commutation of
unbounded self-adjoint operators is a subtle issue (see \cite{rs1}, p. 270-276). Fortunately we shall
not encounter a difficulty of this kind here. But the definition of (generalized) eigenfunctions has
to be clarified. In fact, the rather involved notion of spectral resolution associated to
the electronic Hamiltonian (see the Appendix) is hidden behind this.\\
Let us start with a simple situation. We consider the multiplication operator $T$ by the real variable
$t$ in $\rL^2(\R_t)$ (this is a toy model for the position operators studied in \cite{w}). One can
verify that $T$ is self-adjoint and compute its spectral resolution $S$. The latter is the identity
operator on $\rL^2(\R_t)$ times the Lebesque measure $dt$ on $\R$ (that is the measure that associates
to any finite interval its length). If $\Omega$ is such an interval and $\tau\in\rL^2(\R_t)$ then
$S(\Omega)\tau =|\Omega|\tau$, where $|\Omega|$ is the length of $\Omega$. The completeness of $S$
is expressed by the formula
\[\tau \ =\ \int_\R\, dS(\lambda)\tau\comma\]
which corresponds to \eqref{eq:superposition} and to \eqref{eq:decomp-phi}. The vectors in the
range of $S(\Omega)$ for all possible $\Omega$ play the r\^ole of the electronic ``eigenvectors''.
There is a kind of orthogonality. If $\Omega$ and $\Omega '$ are disjoint subsets in $\R$ then,
for any $\tau, \tau '\in\rL^2(\R_t)$, it turns out that 
\begin{equation}\label{eq:weak-ortho}
\langle S(\Omega)\tau\, ,\, S(\Omega ')\tau '\rangle \ =\ 0\period
\end{equation}
For any fixed $\lambda\in\R$, $S(\{\lambda\})=0$ since the length of the interval $\{\lambda\}$ is zero.
This means that $\lambda$ is not an eigenvalue of $T$. In this framework,
there is no state with exact position $\lambda$ and can be interpreted by saying that the classical
notion of exact position is not well-suited in a quantum situation. One can try to seek an ``eigenvector''
outside the Hilbert space $\rL^2(\R_t)$. One finds a Dirac {\em distribution} at $\lambda$, denoted
by $\delta (t-\lambda)$. This is not a function! The product of two distributions is not defined
in general. In particular, the square of a Dirac distribution and therefore
\[\langle \delta (t-\lambda)\, ,\, \delta (t-\lambda)\rangle\ =\ \int_\R\delta (t-\lambda)^2\, dt\]
are not defined. So what is the meaning of Equations (3.2.14) and (5.6.8) in \cite{w}? Thus the
introduction of states with definite electronic and nuclear positions in \cite{w} does require a more precise definition.\\
Let us come back to the physical version of the Born-Oppenheimer approximation. For each nuclear
position $x$, $Q(x)$ is a self-adjoint operator. Therefore it has a spectral resolution $S_x$ supported
in its spectrum $\sigma (Q(x))$ (see the Appendix). The latter can be split into two parts: a discrete one and above a continuous
one. The discrete part consists of eigenvalues of finite multiplicity. The continuous part $\sigma _c(Q(x))$ is an interval
$[t_x; +\infty [$ where $t_x:=\inf \sigma _c(Q(x))\leq 0$. The operator $Q(x)$ have no positive eigenvalues but
may have eigenvalues in $[t_x; 0]$ (see \cite{cfks}). The latter are called embedded eigenvalues (in the continuous spectrum).
The closed interval $[t_x; 0]$ also contains so-called ``thresholds''. At such a point the nature of the spectrum changes.
For instance, $t_x$ is a threshold which separates discrete and continuous spectra. $0$ is also a threshold at which
a change of ``multiplicity'' of the spectrum occurs. One can have such thresholds between $t_x$ and $0$. 
The number of eigenvalues below $t_x$ may be infinite. This is known as the Efimov effect (see \cite{wa2,y}). In this
case, the eigenvalues accumulate at $t_x$. The (possible) embedded eigenvalues have finite multiplicity and can accumulate
only at thresholds (cf. \cite{abg}). This complicated structure is reflected in the spectral measure.
The discrete eigenvalues below $t_x$ produces the first term on the r.h.s of \eqref{eq:decomp-phi}, that
is a sum of the orthogonal projections in the $y$ variables $\langle \psi_{jk}(x; \cdot ), \cdot)
\rangle \psi_{jk}(x; \cdot )$. From the second term, one can extract the contribution of the embedded eigenvalues producing in this
way similar terms to the first one. In the rest, which is an integral over $[t_x; +\infty[$, one can write, outside the set of
thresholds, $dS_x(\lambda )=S_x^d(\lambda )d\lambda$ for some operator-valued function $\lambda\donne S_x^d(\lambda )$ (see \cite{abg}).
The superscript $d$ refers to ``density'' since $S_x^d$ is, away from the thresholds, the density of the measure $dS_x$ w.r.t.
the Lebesgue measure $d\lambda$. It can be interpreted as the derivative w.r.t. $\lambda$ of $\lambda\donne S_x(]-\infty ; \lambda ])$.
Denote by $\cT _x$ the set of thresholds of $Q(x)$. One can split the set of embedded eigenvalues
of $Q(x)$ in a disjoint union of sets $\cA _x(t)$ for $t\in \cT _x$, such that the elements of each $\cA _x(t)$
can only accumulate at $t$. For $\lambda\in\cA _x(t)$ denote by $k(\lambda )$ its multiplicity and
by $\psi _{\lambda ; 1}(x; \cdot ), \cdots , \psi _{\lambda ; k(\lambda)}(x; \cdot )$ a set of normalized eigenvectors
associated to $\lambda$. By the above computation we may rewrite \eqref{eq:decomp-phi} as 
\begin{eqnarray}\label{eq:decomp-phi-bis}
\varphi (x, \cdot )&=&\sum_{j=1}^J\sum _{k=1}^{k_j}\langle \psi _{jk}(x; \cdot ), \varphi (x; \cdot)\rangle _y
\psi _{jk}(x; \cdot )\nonumber\\
&&\, +\, \sum_{t\in\cT _x}\sum_{\lambda\in\cA _x(t)}\sum _{k=1}^{k(\lambda)}\langle \psi _{\lambda ; k}(x; \cdot ), \varphi (x; \cdot)\rangle _y
\psi _{\lambda ; k}(x; \cdot )\\
&&\, +\, \int _{\lambda\geq t_x}S_x^d(\lambda )\varphi (x; \cdot)\, d\lambda
\period\nonumber
\end{eqnarray}
Note that in the integral in \eqref{eq:decomp-phi-bis}, the integrand is defined outside the discrete set $\cT _x$. 
\eqref{eq:decomp-phi-bis}, as well as \eqref{eq:decomp-phi}, is a correct superposition of electronic states (for a fixed nuclear position at $x$). \\
In the special case where $Q(x)$ is a one-electron Hamiltonian, $\cT _x=\{0\}$ and one expects to be able to
construct generalized eigenfunctions in the continuum in order to express the operators $S_x^d(\lambda )$.
This has been done in a similar situation (see \cite{rs3}, p. 96). Thus we would get a justified version of \eqref{eq:superposition}.
Note that the indices $a$ in the latter such that $\lambda _a=0$ would be absent, if $0$ is not an eigenvalue, and would contribute to
the second term on the r.h.s. of \eqref{eq:decomp-phi-bis}, if $0$ is an eigenvalue. Even in this simple case, we do not know the
behaviour of $S_x^d$ near $0$. There exist some results on this question (see \cite{jk}) but not exactly for the electronic Hamiltonian
we consider here. For a many-body Hamiltonian $Q(x)$ we a priori have the full complexity of \eqref{eq:decomp-phi-bis} and again
the behaviour of $S_x^d$ near $\cT _x$ is not known. See however \cite{wa1} in a similar situation. Furthermore we are not aware of
a construction of generalized eigenfunctions in the continuum for $N$-body systems. \\
To get a rigorous version of \eqref{eq:coupled}, we could try to let $H-E$ act on \eqref{eq:decomp-phi-bis}.
To this end we would essentially need to differentiate w.r.t. $x$ many terms in \eqref{eq:decomp-phi-bis} like $t_x$, $S_x^d(\lambda )$ but also
the sets $\cT _x$ and $\cA _x(t)$. This would probably be a quite involved task, if it is at all possible in some sense. Assume it is.
Then we can take the scalar product in the $y$ variables with $\psi _{jk}(x; \cdot)$ and probably get an equation for
$\theta _{jk}(x)=\langle \psi _{jk}(x; \cdot ), \varphi (x; \cdot)\rangle _y$ similar to \eqref{eq:coupled}, using the fact that
$\psi _{jk}(x; \cdot)$ is orthogonal to the last two terms in \eqref{eq:decomp-phi-bis}. This should work also if we take the scalar
product with $\psi _{\lambda ; k}(x; \cdot )$ for the same reason. In both cases, the terms corresponding to the $c_{a, b}$ in
\eqref{eq:coupled} would be more complicated. Now, to complete the analogy with \eqref{eq:coupled}, we would seek an equation for
$S_x^d(\lambda )\varphi (x; \cdot)$ or perhaps $S_x^d(\lambda )$, for a fixed value $\lambda\geq t_x$. To our best knowledge,
it is however not known if the range of $S_x^d(\lambda )$ is orthogonal to the one of $S_x^d(\lambda ')$ if $\lambda$ and $\lambda '$
are different. We do have here a weak form of orthogonality similar to \eqref{eq:weak-ortho} but it is not clear that it is sufficient
to extract an equation for $S_x^d(\lambda )\varphi (x; \cdot)$ or $S_x^d(\lambda )$ for fixed $\lambda$. In the above strategy, we
neglected the difficulty of the control of $S_x^d(\lambda )$ and its $x$-derivatives up to second order near the thresholds. 
But the behaviour of $S_x^d$ at thresholds is already an open problem.\\
To summarize our discussion on Equations~\eqref{eq:superposition} and~\eqref{eq:coupled}, we have seen that it is possible to produce a rigorous
superposition of electronic states at fixed nuclear positions, namely \eqref{eq:decomp-phi} or \eqref{eq:decomp-phi-bis}. If one prefers a
version in the full $\rL^2$ space, one can use the formula
\begin{equation}\label{eq:direct-int-phi}
\varphi\ =\ \int^{\oplus}\varphi (x; \cdot )\, dx\comma
\end{equation}
where each $\varphi (x; \cdot )$ is given by \eqref{eq:decomp-phi} or \eqref{eq:decomp-phi-bis} (cf. the Appendix).
Equation~\eqref{eq:direct-int-phi} is a well-defined, correct superposition of ``electronic states'' that can replace
\eqref{eq:superposition}. Actually \eqref{eq:direct-int-phi} may be interpreted as the representation of $\varphi$ in
a ``basis'' of common ``eigenvectors'' of $Q$ and of the nuclear position operators. Along this lines, the derivation
of a precise version of \eqref{eq:coupled} seems to be quite difficult, much more difficult than the usual, simple (formal) derivation
of \eqref{eq:coupled}. Thus, if there is another, more clever
construction of electronic ``eigenvectors'' for the continuous spectrum that leads to a well-defined version
of~\eqref{eq:superposition} and~\eqref{eq:coupled}, it deserves to be explained in detail.\\
If one gives up the exact computation of the wave function $\varphi$, one still have the opportunity to use the previous ideas
to look for a good approximation of it. That is precisely what we did in Section~\ref{math-view} with the help of an energy argument.
The latter allowed us to replace \eqref{eq:superposition} by the approximation \eqref{eq:approx} (see also \eqref{eq:born-huang-form}),
that only contains well-defined, quite elementary terms. The controversial Equation~\eqref{eq:coupled} is then replaced by easily
derived coupled differential equations on the functions $\theta _{jk}$ (see just after \eqref{eq:born-huang-form}). In particular,
we were able to forget about the contribution of the continuous spectrum of the electronic Hamiltonians $Q(x)$.

In the framework of Section~\ref{math-view}, ``making the Born-Oppenheimer approximation'' in the
sense used in Physics and Chemistry amounts to removing ``off-diagonal'' terms from $\Pi H\Pi$ (or
$\Pi H_0\Pi$), that is to removing terms involving $\psi _{jk}(x; \cdot )$ and $\psi _{j'k'}(x; \cdot )$
for different couples $(j; k)$ and $(j'; k')$.
In other words, the approximation is performed by
deleting in the differential equations for $\theta _{jk}$ all terms involving $\theta _{j'k'}$ for
$(j'; k')\neq (j; k)$. This approximation
may be justified if the electronic levels $\lambda _j$
are simple and well separated (see \cite{me}). This is also the case when the nuclear kinetic
energy is small as in \cite{bo}, as we saw in the first energy argument in Section~\ref{math-view}.
In general, we should expect the breakdown of this approximation. Recall that, when the total energy
belongs to $(E_-;E_+)$ in fig. 1, we have seen that the nuclear kinetic energy is not small and we have eigenvalue crossings, that is a change of multiplicity.
In a time-dependent or scattering situation, we even expect that the assumption of well separated
electronic levels hinders significant inelastic phenomena (see Section~\ref{math-results} below). \\
In contrast to the usual treatment of the Born-Oppenheimer approximation, we fixed in
Section~\ref{math-view}, at the very beginning, the total energy of the system, selected
certain electronic levels (the $\lambda _j$) ``relevant'' for this energy, and then constructed
the adiabatic Hamiltonian that should approximate $H$ near this energy. In the
first energy argument, when the total energy is close to the infimum of $\lambda _1$, we have
roughly shown that the nuclear kinetic energy is small and recovered one result in \cite{bo},
without using the assumption that the nuclei stay close to some position where the infimum of
$\lambda _1$ is attained. In fact, this property can be recovered from the analysis in \cite{kmsw}
(see comments in Section~\ref{conclusion}). It is again an energy argument, namely the bound
$\Pi ^\perp H_0\Pi ^\perp\geq E_+$ on the range of $\Pi ^\perp$, that allows us to show
(in the limit $\epsilon\to 0$), that the component of $\varphi$ in the range of $\Pi^\perp$ is
negligible to first order, that is the high energy electronic levels have a small contribution to the
superposition \eqref{eq:decomp-phi} of ``electronic'' states that reproduces $\varphi$.
In Section~\ref{math-view} we actually used \eqref{eq:decomp-phi} in the form of \eqref{eq:phi-Pi}.
In particular we can recover the main
issues of the Born-Oppenheimer approximation without relying on the complicated notion
of spectral resolution. \\
The bound $\Pi ^\perp H_0\Pi ^\perp\geq E_+$ on the range of $\Pi ^\perp$ is probably the
core of the mathematical Born-Oppenheimer approximation and plays a rôle in all mathematical
results, that we shall review in Section~\ref{math-results} below. Note that this bound
crucially relies on the fact that the nuclear kinetic energy is bounded below. If it would be 
described by a Hamiltonian that is unbounded below (like a Dirac operator), we would have
to expect that very high energy electronic levels do play a rôle, even if the total energy
of the system is fixed.

\section{Review of mathematical results.}
\label{math-results}
\setcounter{equation}{0}

In this section, we present some rigorously proved results on the Born-Oppenheimer
approximation that illustrate the main ideas developed in Section~\ref{math-view}.
As announced in the Introduction, we decided to set out results, the proof of which
relies on sophisticated mathematical notions and methods. Even the statement of these
results might be difficult to understand. So we try to give an intuitive account of them. 
Since we cannot review all results, we selected one from each of the following
fields: bound states, resonances, scattering process (collision), and time evolution.
These choices may be detected as arbitrary (they reflect the way the author senses the
subject) but we try to present results with the highest degree of generality.
Nevertheless we also comment on other results in these fields. At the end of the
present section, we add some remarks when symmetries of the particles are taken into
account. 

Let us begin with the study of bound states of a molecule which was performed in the
paper \cite{kmsw} (previous results were obtained in \cite{ha3,ha4}). One studies the eigenvalues
of the operator $H$ (cf. \eqref{eq:def-H}) in the framework introduced in
Section~\ref{math-view}. In particular, the adiabatic operator $\Pi H\Pi$ is used first as an effective
Hamiltonian but in a slightly different way. The authors use a so-called Grushin problem and
pseudodifferential techniques to produce a more accurate effective Hamiltonian $F(E)$ (depending
on the sought after energy $E\in (E_-; E_+)$), which is a pseudodifferential matrix operator. $F(E)$
essentially corresponds to the operator that defines the (a priori) coupled equations on the
$\theta _{jk}$ we mentioned at the end of Section~\ref{math-view}. Then $E$ is an eigenvalue of
$H$ (essentially) if and only if $0$ is an eigenvalue of $F(E)$ (cf. Theorem 2.1). Here we mean that,
if $E$ is a true eigenvalue of $H$, then $0$ is an eigenvalue of $F(E')$ where $|E-E'|=O(\epsilon^N)$,
for all integer $N$, and also that, if $0$ is an eigenvalue of $F(E')$ then $H$ has an eigenvalue
$E$ such that $|E-E'|=O(\epsilon^N)$, for all integer $N$. An explicit but rather complicated, infinite
construction produces the operators $F(E)$. For practical purpose, one follows only an appropriate
finite number of steps of this construction to get an operator $F_p(E)$ such that the above errors
are $O(\epsilon^p)$. A concrete example is given in Proposition 1.5, where eigenvalues of $H$ in some
particular energy range are computed up to $O(\epsilon^{5/2})$. \\
For diatomic molecules, the authors
consider an energy range close to the infimum of $\lambda _1$ (like $(E_-^0; E_+^0)$ in fig. 1).
Recall that, for all nuclear positions $x$, $\lambda _1(x)$ is the lowest eigenvalue of the
electronic Hamiltonian $Q(x)$,
which is simple. Actually, one does not need to consider the lowest eigenvalue but it is important
that it is simple and that the rank of the projection $\Pi (x)$ is always $1$. In the mentioned
energy range, the eigenvalues of $H$ and the corresponding eigenvectors are computed by an asymptotic
expansion in power of $\epsilon^{1/2}$ of WKB type. In particular, the original result of \cite{bo} for diatomic
molecules is contained in Theorem 3.2 in \cite{kmsw}. \\
For polyatomic molecules, the same situation is studied
but two cases occur. Recall that $\Gamma$ denotes the set of nuclear positions
$x=(x_1, \cdots , x_{M-1})\in\R^{3(M-1)}$ where the infimum of $\lambda _1$ is attained.
It is assumed that $\Gamma$ is the set of all points $(Ox_1^0, \cdots , Ox_{M-1}^0)$ where $O$
ranges in the set of all orthogonal linear transformations in $\R^3$ and
$x_0=(x_1^0, \cdots , x_{M-1}^0)\in\Gamma$. One can check if the points
$x_1^0, \cdots , x_{M-1}^0\in\R^3$ lie on a line, or on a plane, or generates the whole space
$\R^3$. The molecule is ``linear'', ``planar'', and ``non-planar'' respectively. For a linear or
planar molecule, it is shown that, in an appropriate neighbourhood of $\lambda _1$'s infimum,
there is exactly one eigenvalue of $H$ which is given by a complete asymptotic expansion in
$\epsilon^{1/2}$. A corresponding eigenvector can also be obtained by such an asymptotic expansion. The distance
from this eigenvalue to the rest of the spectrum of $H$ is of order $\epsilon^{5/2}$. In the non-planar
case, two different simple eigenvalues of $H$ are present in the mentioned neighbourhood. The
splitting (that is the distance between these two eigenvalues) is exponentially small in $\epsilon$.
The eigenvalues and the corresponding eigenvectors are again given by an asymptotic expansion
in power of $\epsilon^{1/2}$. These eigenvectors can be related to one another with the help of the reflection
$\varphi (x, y)\mapsto \varphi (-x, -y)$. \\
In the above framework, we mention a modification of the Born-Oppenheimer approximation
performed in \cite{hj6,hj7,hj8} in order to make apparent chemical hydrogen bonds in molecules.
The main idea is to take the hydrogen mass of order $\epsilon^{-3/2}$, while the mass of the heavier atom
and the electronic mass are still of order $\epsilon^{-2}$ and $\epsilon^0=1$, respectively. In this setting,
one can reduce the eigenvalue problem to an effective one in a similar way as in
Section~\ref{math-view}. However, the authors use a multiscale analysis as in \cite{ha3,ha4}. \\
Next we describe the paper \cite{mm} on the resonances of the operator $H$ in the diatomic case.
As we shall see, resonances are complex eigenvalues of an appropriate distortion of $H$. It is
believed that they give information for the long time evolution of the molecules (scattering).
This link has been done in other (simpler) situations but, to our best knowledge, not in the present
framework, i.e. for molecules in the large nuclear masses limit. Actually the authors consider
the diatomic version of the operator $H_0$ (cf. \eqref{eq:def-H_0}), that is the relevant molecular
Hamiltonian without the Hughes-Eckart term. To explain the announced distortion, we need some notation.
Let $\omega :\R^3\to\R^3$ a smooth function, which is $0$ near $0$ and equals the identity map
(i.e. $\omega (x)=x$) for $|x|$ large. For real numbers $\mu$, one introduces the transformation
$\cU _\mu$ defined on total wavefunctions $\varphi (x, y)$ by
\begin{equation*}
(\cU _\mu\varphi )(x, y)\ =\ |J_\mu (x, y)|^{1/2}\, \varphi (x+\mu\omega (x), y_1+\mu\omega (y_1),
\cdots , y_N+\mu\omega (y_N))\comma
\end{equation*}
where the function $J_\mu$ is the Jacobian of the change of variables $(x, y)\mapsto
(x+\mu\omega (x), y_1+\mu\omega (y_1), \cdots , y_N+\mu\omega (y_N))$. It turns out that one
can extend $\cU _\mu$ to small enough complex values of $\mu$. The distorted Hamiltonian is given
by $H_\mu=\cU _\mu H_0\cU _\mu^{-1}$. Since $\cU _\mu$ is unitary for real $\mu$, $H_\mu$ has
the same spectrum as $H_0$ on the real line but, for nonreal $\mu$, the continuous part of
the spectrum of $H_\mu$ is obtained from the one of $H_0$ by some rotation in the complex plane. Furthermore, between
the continuous spectra of $H_0$ and $H_\mu$, nonreal eigenvalues of $H_\mu$ of finite multiplicity
appear. They actually do not depend on $\mu$ and are called the resonances of $H_0$. They
are close to the continuous spectrum of $H_0$, which is responsible for the scattering processes
governed by $H_0$. One wants to compute these resonances. To this end, one faces an eigenvalue
problem as above in \cite{kmsw} but now for the non self-adjoint operator $H_\mu$. In \cite{mm}, it is shown that
one can adapt the arguments of \cite{kmsw} to show that $E$ is a resonance of a modified
version of $H_0$ if and only if $0$ is an eigenvalue of an $E$-dependent pseudodifferential
matrix operator. Due to a technical difficulty, the authors have to smooth out the repulsive,
nuclear interaction, leading to the modified
version of $H_0$. Thus it is still open whether a similar result holds true for the resonances
of $H_0$. Anyhow, the imaginary part of these resonances is expected to be of order $e^{-c/\epsilon}$
in $\epsilon$ with $c>0$ (as shown in \cite{ma2,ma3} in a simpler framework). Since the inverse of the
imaginary part of a resonance is interpreted as the lifetime of the corresponding resonant state, the
resonant states in the present situation should live on a time scale of order $e^{c/\epsilon}$.
Thus such a state could be mistaken for a true bound state and should perhaps be taken into account in
Chemistry, when ``stable'' molecular structures are sought. 

Next we are interested in the non-resonant scattering (or collision) theory for diatomic molecules.
We use the framework presented in Section~\ref{math-view} for the
operator $H'$ (cf. \eqref{eq:H'}) but we impose a stronger decay on the pair interactions.
We replace at spatial infinity the Coulomb interaction $|\cdot |^{-1}$, which has long range, by a short range
potential (essentially of the type $|\cdot |^{-1-\delta}$, for some $\delta >0$).
To present the result in \cite{kmw2}, we need a short review of the short range scattering
theory (see details in Section XI.5 p. 75 in \cite{rs3}).\\
Recall that the free dynamics for large negative time is generated by $H_c'$ in \eqref{eq:H_c'},
which can be written as 
\begin{equation*}
H_c'\ =\ -\frac{1}{2}\Bigl(\frac{1}{M_1}+\frac{1}{M_2}\Bigr)\Delta _x\, +\, T_{\rm HE}'\, +\,
Q^c\period
\end{equation*}
Here we denote by $T_{\rm HE}'+Q^c$ the operator on the $y$ variables and also the same operator acting in the
$x$ and $y$ variables (the latter being in fact the direct integral over the $x$ variables of the former). 
We choose an eigenvalue $E^c$ and a corresponding eigenvector $\psi _c$ of $T_{\rm HE}'+Q^c$ and
consider the scattering process that starts for $t\to -\infty$ by the free motion given by 
\begin{equation*}
-\frac{1}{2}\Bigl(\frac{1}{M_1}+\frac{1}{M_2}\Bigr)\Delta _x
\end{equation*}
of two clusters that are in the state $\psi _c(y)$. The initial state is described by a
wavefunction $\tau (x)\psi _c(y)$, where $\tau$ is a nuclear wavefunction and $\psi _c(y)$ is
actually the (tensor) product of an electronic wavefunction of the cluster $c_1$ by one of the cluster $c_2$.
If we forget about the Hughes-Eckart contribution $T_{\rm HE}'$, these latter wavefunctions represent
respectively an electronic bound state in the atom/ion $c_1$ and another in the atom/ion $c_2$. It
turns out that we can find a wavefunction $\varphi _+(x, y)$ such that, for $t\to -\infty$,
the real evolution of $\varphi _+(x, y)$ is close to the free evolution of $\tau (x)\psi _c (y)$,
that is
\begin{equation*}
\bigl\|e^{-itH'}\varphi _+\, -\, e^{-itH_c'}\tau (x)\psi _c (y)\bigr\|\ \to\ 0\period
\end{equation*}
We denote by $\Omega _+$ the operator $\tau (x)\psi _c (y)\mapsto\varphi _+$. Similarly,
the final state is described by $\tau '(x')\psi _d '(y')$ corresponding to some cluster decomposition
$d$ (with a priori different coordinates) and one can find a wavefunction $\varphi _-(x', y')$
such that, for $t\to +\infty$,
\begin{equation*}
\bigl\|e^{-itH'}\varphi _-\, -\, e^{-itH_d'}\tau '(x')\psi _d' (y')\bigr\|\ \to\ 0\period
\end{equation*}
We define $\Omega _-(\tau '(x')\psi _d' (y'))=\varphi _-$ and we can check that
$\tau '(x')\psi _d'(y')$ can be recovered from $\varphi _-(x', y')$ by application of the
adjoint $\Omega _-^\ast$ of $\Omega _-$. Thus the operator $S:=\Omega _-^\ast\Omega _+$ sends
the initial state to the final state of the scattering process. It is the
scattering operator for this process while $\Omega _+$ and $\Omega _-$ are the wave operators.
The strange sign convention for the wave operators can be interpreted in the following way:
$e^{-itH'}\Omega _+\tau (x)\psi _c (y)$ represents the future (+) evolution for the interactive
dynamics (defined by $H'$) of the free state $\tau (x)\psi _c (y)$. When $c=d$, $\tau =\tau '$,
and $\psi _c =\psi _d'$, we have an elastic scattering process. When $c=d$ but $\psi _c$ and
$\psi _d'$ are orthogonal, the inelastic process corresponds to a change of electronic level
in some cluster (an excitation of an electron in $c_1$ for instance). When $c\neq d$ but
$c_1$ and $d_1$ contain the same nucleus and so do $c_2$ and $d_2$, an electron at least has
moved from one nucleus to the other. We can also consider the case where $c$ is as above
while $d_1$ contains the two nuclei and $d_2$ only electrons (for instance,
two ions forming a molecule and loosing some electrons). Among the inelastic processes we just
described, the two last ones might be interesting for Chemistry. The above construction
of a scattering operator can be done for all possible cluster decompositions $c$ and $d$ and
the collection of the $S$ operators completely describes the possible scattering processes. The same
construction can also be performed for molecules with more than $2$ nuclei with a richer family of
processes of chemical interest. \\
We point out that a scattering theory exists for long range interaction (like the Coulomb one).
Essentially, one has to modify the construction of the wave operators, which become technically
more involved. \\
We come back to the situation studied in \cite{kmw2}, that is for a diatomic molecule with $d=c$
given as in Section~\ref{model} but with short range interactions. We choose an energy range
$(E_-^c, E_+^c)$ as in fig. 2. In particular, it is above the infimum of the spectrum of $H_c'$
in \eqref{eq:H_c'} thus, by the HVZ Theorem (see Theorem XIII.17 p. 121 in \cite{rs4}), this
energy range is included in the continuous part of $H'$ but might contain eigenvalues.
We focus on scattering processes with total energy $E\in (E_-^c, E_+^c)$. In view of 
Section~\ref{math-view}, we replace $E_+$ by $E_+^c$ and consider eigenvalues $\lambda _j(x, \epsilon)$
of $Q_c(x)+T_{\rm HE}'$ that are somewhere less or equal to $E_+^c$. One then constructs the
associated projections $\Pi (x, \epsilon)$ and consider the adiabatic operator $\Pi H'\Pi$.
Let $\Omega ^c_\pm$ be wave operators associated to the decomposition $c$ as above such that
the electronic energy $E^c$ of the initial state satisfies $E^c<E_-^c$ ($E^c$ is close to some
eigenvalue $E^c(0)$ of $Q^c$, see fig. 2). Now one
can also compare the dynamics of $\Pi H'\Pi$ and the free dynamics generated by $H_c'$ and
construct so-called adiabatic wave operators $\Omega ^{\rm AD}_\pm$. The goal is to show that
\begin{equation}\label{eq:approx-wave-op}
\|\Omega ^c_\pm\, -\, \Omega ^{\rm AD}_\pm\|\ =\ O(\epsilon)\comma
\end{equation}
when the wave operators act on wavefunctions with energy in $(E_-^c, E_+^c)$. To this end,
one needs an important assumption, the non-trapping condition, on the classical mechanics
generated by the nuclear, classical Hamilton functions $h_j(q, p)=\|p\|^2+\lambda _j(q, 0)$ (with $\epsilon =0$
and for the above selected eigenvalues $\lambda _j$) at energies in $(E_-^c, E_+^c)$. This
non-trapping condition says that all classical trajectories of energy $E\in (E_-^c, E_+^c)$
for any Hamilton function $h_j$ go to spatial infinity in both time directions. It implies the
absence of eigenvalues in $(E_-^c, E_+^c)$ and prevents resonance phenomena.
In fig. 2, this assumption is satisfied. \\
Under the additional assumption that only the simple eigenvalue $\lambda _1$ is somewhere
less or equal to $E_+^c$ (in particular the image of $\Pi (x, \epsilon )$ is always of dimension $1$),
the approximation \eqref{eq:approx-wave-op} is proved in \cite{kmw2}. An important step in
the proof is to establish an appropriate estimate on the resolvents $\C\setminus\R\ni z
\mapsto (z-H')^{-1}$ and $\C\setminus\R\ni z\mapsto (z-\Pi H'\Pi)^{-1}$ of $H'$ and $\Pi H'\Pi$
respectively and this is done for long range interactions (in particular for the Coulomb one).
Because of the additional assumption, only elastic scattering is covered. In this framework,
we mention the papers \cite{jec2} on the scattering operator and \cite{jkw} on scattering
cross-sections. \\
If one removes the above additional assumption, one can obtain the approximation
\eqref{eq:approx-wave-op} but under the condition that the eigenvalues $\lambda _j$ do not
cross (see \cite{jec1}). In this situation, a similar approximation holds true for
scattering cross-sections and it can be shown that the inelastic scattering is
disadvantaged compared to the elastic one (see \cite{jec3}). In the simplified framework
of a Schrödinger operator with matrix potential, it is even shown in \cite{bm} that the
inelastic scattering is exponentially small in $\epsilon$. Therefore, to study it, we probably have to
accept eigenvalues crossings and we need to control their effect on the scattering.
As mentioned before, the projection $\Pi$ is still smooth but the eigenvalues
$\lambda _j$ might be only continuous and the corresponding eigenvectors $\psi _{jk}$ might be
discontinuous at the crossing. In this situation, we mention the work by \cite{fr} on 
Schrödinger operators with matrix potential and for a special case
of crossing (crossing at just one point), where the resolvent estimates mentioned above are
derived. For some types of crossing, the $\lambda _j$ and $\psi _{jk}$ are smooth and one can
prove the same result (see \cite{jec4,dfj}). This is the case for diatomic molecules thanks to
their radial symmetry with respect to the $x$ variable. In the work in
progress \cite{js}, one uses this property to get the resolvent estimates and also the
approximation \eqref{eq:approx-wave-op} for diatomic molecules.

Now we come to the last field we wanted to consider, namely the Born-Oppenheimer approximation
for the time evolution of molecules, and present results obtained in \cite{ms2}. We consider again
the operator $H$ in \eqref{eq:def-H} but we look for an approximation of the evolution operator
$e^{-itH/\epsilon}$ (i.e. the molecular evolution on a time scale $1/\epsilon$). As in Section~\ref{math-view},
the authors choose a certain energy range (like $(E_-; E_+)$ in fig. 1) and construct
a better projection $\Pi _r$, starting from the operator $\Pi$ adapted to this energy range.
The estimate of the commutator $[H, \Pi]=O(\epsilon)$
is improved in this way in the estimate $[H, \Pi _r]=O(\epsilon^p)$, for all integer $p$. With the help of
$\Pi _r$, the authors introduce a map $\cW$ that transforms wavefunctions $\varphi (x, y)$
for the full molecule into wavefunctions in $x$ only but with values in the $L$-dimensional
vectors ($L$ being the constant dimension of the image of the $\Pi (x)$). This map replaces the
electronic wavefunctions, that live in an infinite dimensional space, by a finite number of
degrees of freedom, namely the coordinates of the $L$-dimensional vectors. We point out here
that no restriction on the number of nuclei is required and that eigenvalue crossings are
allowed. There exists a $L\times L$-matrix operator $A$ acting on the range of $\cW$ such that,
for a large class of initial states $\varphi$ with energy in the chosen energy range, for all integer $p$,
the time evolution of $\varphi$ is given by
\begin{equation}\label{eq:approx-time-evo}
e^{-itH/\epsilon}\varphi\ =\ \cW^\ast e^{-itA/\epsilon}\cW\varphi\ +\, O((1+|t|)\epsilon^p)\comma
\end{equation}
where $t$ ranges in some bounded, $p$- and $\epsilon$-independent interval. So, to compute a good
approximation of the time evolution of $\varphi$, one first lets $\cW$ act, then follow the
evolution of $\cW\varphi$ generated by $A$ (a simpler evolution) and then lets the
adjoint of $\cW$ act. The operator $A$ is obtained by an infinite but explicit construction. If one
accepts an error of size $O((1+|t|)\epsilon^p)$, for a fixed $p$, one can replace $A$ by
an operator $A_p$ which is obtained by a finite procedure. \\
As a consequence of the previous approximation, the authors derive for $L=1$ a rather 
precise description of the time evolution of coherent states (which are probably the simplest
states), completing in this way previous results of this type (for instance in \cite{ha1,ha5}). The
assumption $L=1$ prevents eigenvalue crossings. For the time evolution of coherent states,
the effect of eigenvalue crossings was studied in \cite{ha7}. Even for these states, this effect
is complicated in general and another approach was followed by considering so-called avoided
crossings (see \cite{hj1,hj2}). Instead of having a crossing of the electronic eigenvalues
$\lambda _1$ and $\lambda _2$, one assumes that, for some particular nuclear position, the nonzero
difference $\lambda _1-\lambda _2$ is small (with an appropriate size compared to $\epsilon$). This
approach avoids the technical difficulties carried by true crossings but allows inelastic
phenomena (like the transfer of a wave packet from the electronic level $\lambda _1$ to
$\lambda _2$). In a simplified framework (compared to the molecular setting) but for the time
evolution through true eigenvalue crossings, we mention \cite{fg} in a special case where
the $\lambda _j$ and the eigenvectors $\psi _{jk}$ are not smooth and \cite{dfj} where the latter
are smooth. In \cite{fg} a Landau-Zener formula plays an important rôle. In \cite{dfj}, although the
coupling of the smooth crossing eigenvalues vanishes formally at $\epsilon=0$, a coupling effect between them
is proved in a very special situation, that should be unphysical. Finally we quote the paper
\cite{tw} where the Born-Oppenheimer approximation for the time evolution of molecules coupled
to a quantized radiation field is analysed.

We end this section with some comment on the symmetries of particles. First one should take
into account that the electrons are fermions and consider only antisymmetric electronic wavefunctions.
Second, if the molecule contains two identical nuclei for instance, one should
restrict the nuclear wavefunctions to the ones that are symmetric with respect to the exchange
of these two nuclei. In principle, such constraints can be included in a mathematical
framework but, in practice, this has not been done. Let us give some explanation for this.
Including these symmetries amounts to letting act the operators on smaller Hilbert spaces.
So if one can perform the approximation in the full Hilbert space, it is also valid on a
smaller one. However, the electronic symmetry could change the spectrum of the electronic
Hamiltonian (the eigenvalue $\lambda _3$ could be absent or its multiplicity could be lowered)
but this would change essentially the input of the above mathematical treatment and not the core
of the approximation. Taking into account the nuclear symmetry could give finer results but
this would be hidden in the properties of the adiabatic operator
derived by the mathematical Born-Oppenheimer approximation. Up to now, it seems that there was
no clear motivation from the mathematical point of view to include symmetries; thus it was natural to
avoid them and the technical complications they carry.

\section{Conclusion.}
\label{conclusion}
\setcounter{equation}{0}

During the review of the traditional Born-Oppenheimer approximation (in Section~\ref{traditional-bo}), we have pointed out
some imprecisions that can be partially cured with the help of a sophisticated mathematical tool (see Section~\ref{comparison}).
In particular we have explained the mathematical difficulties to derive (a precise version of) important coupled differential equations,
that are claimed to be exact in the usual, physical presentation of the theory. 
We have presented the way mathematicians consider the approximation and emphasise the following main
difference. In the mathematical point of view, one fixes first the total energy of the molecular system
and then constructs an effective Hamiltonian depending on this energy. The accuracy of the Born-Oppenheimer
approximation is then defined by the quality of the replacement of the true Hamiltonian by the effective one
and is measured in terms of a small parameter related to the electron/nucleon mass ratio, enlarging in this
way the traditional definition of the sentence ``making the Born-Oppenheimer approximation''. 
This procedure turns out to be universal in the sense that it applies to different physical situations
(bound states, time-evolution, resonances, non-resonant scattering).
We also have seen that the mathematical approach actually provides a presentation of the approximation
that avoids the mentioned, complicated tool and relies on more elementary arguments and well-defined objects.\\
We have presented the essential structure of the mathematical justification of the Born-Oppenheimer
approximation and tried to illustrate it on concrete results on bound states, on the time evolution,
and in scattering theory. In particular, we have seen that the basic idea, namely the use of electronic
levels and eigenfunctions, is actually the same as in \cite{bh,me,sw,w} but with an important additional feature:
the use of an appropriate projection onto a finite number of electronic states. One writes the full Hamiltonian
as the sum of the nuclear kinetic energy, of an electronic Hamiltonian, and of comparatively smaller terms,
mimicking in this way the usual framework for the well-developed semiclassical analysis. Indeed, taking the favorite example
of this analysis, namely the semiclassical Schrödinger operator $-\epsilon^2\Delta _x +V(x)$, the
nuclear kinetic energy stands for the semiclassical Laplace operator $-\epsilon^2\Delta _x$ while the
electronic Hamiltonian plays the rôle of the potential $V$. Note that a semiclassical touch is already
present in the original work \cite{bo} when expansions around a
 nuclear equilibrium positions are performed.
We have explained how the full
Hamiltonian can be approximated by a so-called adiabatic operator, the construction of which
essentially rests upon the electronic Hamiltonian (or clamped-nuclei Hamiltonian). Even
the construction of the refined projection $\Pi _r$, which leads to a very accurate approximation,
completely depends on this Hamiltonian. We point out that our intuitive argument to compute
an eigenvalue and an eigenvector of the full Hamiltonian (the operator $H$), up to an error
$O(\epsilon)$, actually leads to a modification of Born-Huang's proposition of approximated eigenvalue and eigenvector
(see \eqref{eq:born-huang-form}), the change consisting in a limitation of the number of electronic levels taken into
account. This modified Born-Huang's approach is legitimate but not very accurate. To go beyond, as we mentioned, one needs
to take into account the variation of the electronic Hamiltonian with respect to the nuclear
variables. When we look for an eigenvalue close to the groundstate energy (which is close to
the infimum of the lowest electronic eigenvalue $\lambda _1$), we have seen that the nuclear
kinetic energy is small, as a consequence of this closeness and not of the large size of the
nuclear masses. In particular, the original computation in \cite{bo} is legitimate. The situation
is different for higher energy but it can be handled with the help of semiclassical
analysis (see \cite{kmsw}), as explained in Section~\ref{math-results}.
Concerning the scattering (or collision) theory and the time evolution of molecules, we reviewed
some results and pointed out the main difficulty, namely the control of eigenvalue crossings.
In particular, this difficulty hinders the treatment of chemically relevant situations but
we stressed that some progress was made. Letting $\epsilon$ tend to $0$ instead of keeping its
physical value is essential in all the above mathematical works but, as we noticed, it
might be inappropriate in some physical or chemical situations. \\
In \cite{sw}, the authors subscribed to Löwdin's impression (expressed in \cite{l}), that
it might be difficult to extract from the molecular Hamiltonian the concrete realization of
chemical concepts like isomerism, conformation, chirality. Probably, they are right but
the situation is perhaps not hopeless. We pointed out the papers \cite{hj6,hj7,hj8} that
try to describe hydrogen bonds. In the paper \cite{jkw}, it was proved that some symmetries
in the ion-atom scattering influence the leading term of scattering cross-sections in the
large nuclear masses limit. The techniques used in \cite{kmsw} tells us that, near the minimum
of a nondegenerate electronic eigenvalue $\lambda _j$, one can find a bound state of the
molecule with low nuclear kinetic energy. In this state, the nuclei vibrate near an equilibrium
position, located where the minimum is attained, in the sense that the wave function is concentrated
in the nuclear variables near this position. If one can compute (numerically) this position,
one gets the nuclear structure of this bound state (internuclear distances, symmetries).
Because of computational error, it might be difficult to check if the molecule is planar or
not. By light excitation, one can measure the difference between the molecular energies, that
are the two closest levels to the minimum of $\lambda _j$. If the difference is ``very'' small,
then the molecule in this state is not planar and if the difference is big ``enough'', then
it is planar, thanks to \cite{kmsw}. Of course, these examples are limited from the chemical
point of view but show that simple properties of the molecular structure can be extracted from the
molecular Hamiltonian. We also stress that there exist tools, like the theory of
(co-)representations, to take into account symmetries of molecules. An example
of such use in the molecular context is provided in \cite{ha7}. \\
We emphasise that, in the mathematical treatment of the Born-Oppenheimer approximation, the
nuclei are always considered as quantum particles. The use of clamped nuclei is just a tool
to construct an appropriate effective Hamiltonian but the latter is a quantum, nuclear Hamiltonian
with restricted electronic degrees of freedom. It seems that the Born-Oppenheimer approximation
in Chemistry often reduces to the computation of the molecular potential energy surfaces
and to classical motion of the nuclei on these surfaces.
Except for the special situation studied in \cite{bo} (concerning the ground state), a classical
treatment of the nuclei is not justified, as already pointed out in \cite{sw}. This does not mean
that classical behaviours of the nuclei do not emerge. On the contrary, it is a general fact that
semiclassical situations (like the molecular one with small $\epsilon$) are strongly connected to
classical features and often produce effects that resemble classical ones at the macroscopic
level. For instance, we have seen that the nuclear, classical Hamiltonians $h_j$ play a r\^ole in the
scattering situation and explained above that, under some circumstances, the nuclei can be viewed
as classical particles vibrating near an equilibrium position. A classical use of potential energy
surfaces can be applied to find some molecular excited energy near the bottom of an electronic
potential well but may not capture all of them. So we subscribe to the warning addressed to
Chemists in \cite{sw}. \\
To the presentation of the Born-Oppenheimer approximation in \cite{bo,bh,me,sw,w} we essentially added
the notion of adiabatic Hamiltonian associated to a chosen total energy as a central tool (and removed
some undefined objects and equations). It
is present in all the mathematical results discussed above. For this reason, the mathematicians
consider the Born-Oppenheimer approximation as valid if the true Hamiltonian can be replaced
by the adiabatic one up to some controlled error depending on the semiclassical parameter. This
property could be called an adiabatic reduction or approximation in order to avoid confusion with
the traditional Born-Oppenheimer approximation in Physics. We also insisted on the semiclassical structure
of the mathematical approximation that is close to the typical model $-\epsilon^2\Delta _x +V(x)$. Our
optimistic view of future mathematical developments actually relies on the power and the diversity
of techniques provided by the semiclassical analysis. \\
The actual mathematical treatment of the Born-Oppenheimer approximation for molecular systems
is expressed in a rather involved language and provides a theoretical information on such systems,
that might be considered as unsatisfactory from the physical or chemical point of view. We
tried to make it accessible to a large readership and to show that, despite the real difficulties
it has to face, it could be improved, taking more and more into account physical and chemical
preoccupations.

\section{Appendix}
\label{appendix}
\setcounter{equation}{0}

In this appendix we recall, in a intuitive but not very precise way, some mathematical notions
used in the text. More accurate informations about these notions are provided in \cite{rs1,rs2,rs4,t1,t2}.\\

{\em Infimum and exponential smallness.} The infimum of a subset of $\R$ (the set of real numbers) that is bounded below is
the largest lower bound of this set. If this set is the image of a real valued function,
its infimum defines the infimum of the function. In the cases considered in the text,
the infimum of a function is actually its minimal value. A real or complex valued function $f$ of
a positive $\epsilon$ is exponentially small in $\epsilon$ if there exist $c, C>0$ such that,
for $0<\epsilon\leq 1$, $|f(\epsilon )|\leq Ce^{-c/\epsilon}$. It is indeed small near $0$ since
$Ce^{-c/\epsilon}$ tends to $0$ as $\epsilon\to 0$.

{\em $\rL^2$-spaces.} The squared $\rL^2$-norm $\|f\|^2$ of a complex valued function $f$ is the integral of the modulus squared of the function $f$. If it is finite, we say that the function belong to the space $\rL^2$ which is a normed
vector space. Such a function is normalized if its $\rL^2$-norm equals one. The $\rL^2$-norm is
associated to the scalar product
\[\langle f, g\rangle =\int \overline{f}(t)g(t)\, dt\period\]
In the space $\rL^2$ there are functions
that are not differentiable in the usual sense. It turns out that one can define for them
generalized derivatives (in the sense of distribution). In particular, the wave function $\varphi$
is not everywhere regular in the usual sense but $\Delta _x\varphi$
has a meaning as $\rL^2$ function. We say that a function $0<\epsilon \donne f(\epsilon)\in \rL^2$ is
$O(\epsilon )$ in $\rL^2$-norm near $0$ if there exists $C>0$ such that, for all $0<\epsilon \leq 1$,
$\|f(\epsilon)\|\leq C\epsilon$.

{\em Operators in $\rL^2$-spaces, self-adjointness, spectrum.} In the text we consider operators acting in $\rL^2$. Such an operator $A$ is defined on a ``large enough'' subset $D(A)$ of the space $\rL^2$ and, for $f\in D(A)$, $Af$ is again in $\rL^2$. The range of $A$ is the set of all $Af$ for $f\in D(A)$. It is a vector subspace of $\rL^2$. The rank
of $A$ is the dimension of the range of $A$. An operator $A$ is bounded if there exists $C>0$ such
for all $f\in D(A)$, $\|Af\|\leq C\|f\|$. Such an operator can be extended to the whole $\rL^2$
with the previous property preserved. A projection $P$ of $\rL^2$ is a bounded operator
defined on $\rL^2$ satisfying $P^2=P$ (here $P^2$ denotes the composition of $P$ by $P$).\\
For appropriate operators $A$ in $\rL^2$, one can define an adjoint denoted by $A^\ast$.
It is defined on an appropriate domain $D(A^\ast)$ and satisfies, for $f\in D(A^\ast)$ and $g\in D(A)$,
$\langle A^\ast f, g\rangle = \langle f, Ag\rangle$. Sometimes $A^\ast$ is an extension of $A$,
that is $D(A)$ is contained in $D(A^\ast)$ and $Af=A^\ast f$ for $f\in D(A)$. Such an operator $A$ is symmetric.
Sometimes there exist an extension $B$ of a symmetric $A$ such that $B^\ast=B$ in the sense that
$D(B^\ast)=D(B)$ and $Bf=B^\ast f$ for $f\in D(B)$. $B$ is called self-adjoint. If $A$ admits such
a self-adjoint extension, one says that $A$ has a self-adjoint realization. One can show that
the Laplace operator defined on smooth functions with compact support is symmetric and has a
(unique) self-adjoint realization in the space of $\rL^2$ functions $f$ with
(distributional) $\Delta f$ in $\rL^2$. A self-adjoint projection on $\rL^2$ is called an
orthogonal projection. It turns out that, for a self-adjoint operator $A$,
the operator $z-A$ for nonreal $z$ is invertible and the inverse $(z-A)^{-1}$ is bounded.
The function $z\donne (z-A)^{-1}$ is called the resolvent of $A$. Sometimes it can be extended
to some strict subset of $\R$. The complement of this subset is called the spectrum $\sigma (A)$
of $A$. Further informations can be found in \cite{rs1,rs2}.

{\em Spectral resolution.} A complex measure $\mu$ on $\R$ is a function that associated to appropriate subsets of $\R$ a
complex value. The Lebesgue measure is the one that associates to these subsets their length.
One can define an integral w.r.t. $\mu$: $r\donne \int r(\lambda)d\mu(\lambda)$, where $r(\lambda)$
is an appropriate complex-valued function on $\R$. For the Lebesgue measure,
this integral is the usual integral on $\R$. Now we come to the delicate notion of spectral
resolution of a self-adjoint operator (see \cite{rs1} p. 234). Such a spectral resolution on $\rL^2$ is a projection-valued
measure $S$ on $\R$ satisfying some properties. For appropriate subsets $\Omega$ of $\R$, $S(\Omega )$ is an
orthogonal (self-adjoint) projection on $\rL^2$ and, if $\Omega '$ is disjoint from $\Omega$, then $S(\Omega )S(\Omega ')=S(\Omega ')S(\Omega )=0$. The latter property is actually an orthogonality property. For $f, g\in\rL^2$, $s_{f, g}: \Omega\donne \langle f, S(\Omega )g\rangle$ is a complex measure. The integral over $\R$ of this measure gives $\langle f, g\rangle$. More generaly, if $r$ is an appropriate complex-valued function of a real variable, then one can define
an operator $B$ satisfying, for appropriate $f, g\in\rL^2$,
\[\langle f\, ,\, Bg\rangle \ =\ \int_\R r(\lambda )\, ds_{f, g}(\lambda)\ =:\ \int_\R r(\lambda )
\, d\langle f\, ,\, Sg\rangle (\lambda)\period\]
In this way, one can define an operator-valued integral associated to $S$: $\int r(\lambda)dS(\lambda)=B$.
In particular, $S(\Omega )=\int r(\lambda)dS(\lambda)$, where $r$ is the characteristic function of $\Omega$, and the identity operator $I$ on $\rL^2$ is given by $I=\int 1\, dS(\lambda)$. \\
It turns out that a self-adjoint operator $A$ can be written as $A=\int \lambda dS(\lambda)$ in an essentially
unique way. The corresponding $S$ is called the spectral resolution of $A$. It lives on the spectrum
$\sigma (A)$ of $A$ in the sense that, if $\Omega$ does not intersect $\sigma (A)$, then $S(\Omega )=0$.
If $E$ is an eigenvalue of $A$ (this means that there is a nonzero function $\psi$ in $D(A)$ such that $A\psi =E\psi$) then
$S(\{E\})$ is the projection onto the space $\{\psi \in D(A); A\psi =E\psi\}$. If $E$ is not an eigenvalue,
then $S(\{E\})=0$. The orthogonality property mentioned above implies in particular that, if $A\psi =E\psi$
and $E\not\in \Omega$, then, for any function $f$,
\[\langle \psi\, ,\, S(\Omega)f\rangle \ =\ \langle S(\{E\})\psi\, ,\, S(\Omega)f\rangle\ =\ \langle \psi\, ,\,
S(\{E\})S(\Omega)f\rangle\ =\ 0\period\]
The spectral resolution $S$ realizes a ``diagonalization'' of $A$. 
If $A$ has only the eigenvalues $E_1, \cdots , E_n$ in its spectrum, then $S$ is the sum over $j$ of the Dirac
delta distribution at $E_j$ times the projection $S(\{E_j\})$ onto the associated spectral subspace.
Furthermore $A=\int \lambda dS(\lambda)$ becomes $A=\sum E_jS(\{E_j\})$ in this case, that is a diagonalization
of $A$. \\
If $A$ has the eigenvalues $E_1, \cdots , E_n$, that are below some real $T$, and if $[T; +\infty[$ is the ``continuous spectrum'' of $A$ then 
\[
A\ =\ \sum_{j=1}^n E_jS(\{E_j\})\, +\, \int_T^{+\infty} \lambda dS(\lambda)\hspace{.4cm}\mbox{and}\hspace{.4cm}I\ =\ \sum_{j=1}^n S(\{E_j\})\, +\, \int_T^{+\infty} 1 dS(\lambda)\period \]
This still holds true if $n$ is infinite provided the sums 
\[\sum_{j=1}^\infty E_jS(\{E_j\})\hspace{.4cm}\mbox{and}\hspace{.4cm}\sum_{j=1}^\infty S(\{E_j\})\]
are interpreted in the following way: For all integer $j$, let $T_j$ be a linear operator on $\rL^2$. Let  
$\psi\in\rL^2$. If there exists some $\varphi\in\rL^2$ such that the sequence 
\[\left(\left\|\varphi\, -\, \sum_{j=1}^N T_j\psi \right\|\right)_N\]
tends to zero (the norm $\|\cdot\|$ being the $\rL^2$-norm) then we write 
\[\varphi \ =\ \left(\sum_{j=1}^\infty T_j\right)\psi \ =\ \lim _{N\to\infty}\, \sum_{j=1}^N T_j\psi\period\]
If this holds true for any $\psi\in\rL^2$, this defines a linear operator $\sum_{j=1}^\infty T_j$ on $\rL^2$.

{\em Direct integrals.} Let us explain notions of direct integral associated to the Hilbert space $\rL^2(\R^n_x\times\R^d_y)$
(details in a more general setting can be found in \cite{rs4} p. 280-287). This space may be viewed as the set
$\rL^2(\R^n_x; \rL^2(\R^d_y))$ of square integrable functions of $x\in\R^n$ with valued in the Hilbert space $\rL^2(\R^d_y)$.
We express this by the following direct integral of Hilbert spaces
\begin{equation*}
\rL^2(\R^n_x\times\R^d_y)\ =\ \rL^2(\R^n_x; \rL^2(\R^d_y))\ =\ \int^{\oplus}\rL^2(\R^d_y)\, dx\period
\end{equation*}
A vector $\varphi\in\rL^2(\R^n_x\times\R^d_y)$ can be written as the vector direct integral
\begin{equation}\label{eq:vector-direct-integral}
\varphi\ =\ \int^{\oplus}\varphi (x; \cdot)\, dx\comma
\end{equation}
where $\varphi (x; \cdot)$ denotes the $\rL^2(\R^d_y)$-function $y\donne\varphi (x; y)$. Furthermore,
we recover the $\rL^2(\R^n_x\times\R^d_y)$-norm $\|\varphi \|$ of $\varphi$ by the usual integral 
\begin{equation*}
\|\varphi \|^2\ =\ \int\|\varphi (x; \cdot)\|_y^2\, dx\comma
\end{equation*}
where $\|\cdot \|_y$ denotes the $\rL^2(\R^d_y)$-norm. More generally, the $\rL^2(\R^n_x\times\R^d_y)$
scalar product $\langle \varphi , \psi\rangle$ is given by 
\begin{equation*}
\langle \varphi , \psi\rangle\ =\ \int\langle \varphi (x; \cdot), \psi (x; \cdot)\rangle_y
\, dx\comma
\end{equation*}
where $\langle \cdot , \cdot\rangle_y$ is the $\rL^2(\R^d_y)$ scalar product. This construction seems to be
quite artificial and useless. \eqref{eq:vector-direct-integral} only expresses the fact that one knows
completely $\varphi$ as soon as one knows the functions $y\donne\varphi (x, y)$ for all $x$. However
this construction is used to consider parameter dependent operators. A function $x\donne A(x)$ with
values in the self-adjoint operators in $\rL^2_y:=\rL^2(\R^d_y)$ and with appropriate regularity defines an operator $A$ in 
$\rL^2:=\rL^2(\R^n_x\times\R^d_y)$ by the requirement that, for all $x$ and appropriate $\varphi\in\rL^2$,
$(A\varphi )(x; \cdot)=A(x)\varphi (x; \cdot)$. This means that, for all $x$, the $\rL^2$ function $A\varphi$ at $x$ is an $\rL^2_y$ function defined by the
action of $A(x)$ on the $\rL^2_y$ function $y\donne \varphi (x, y)$. One sets
\begin{equation*}
A\ =\ \int^{\oplus}A(x)\, dx\period
\end{equation*}
For example, the identity operator on $\rL^2(\R^n_x\times\R^d_y)$ is the direct integral of the constant function
$x\donne I$, where $I$ is the identity operator on $\rL^2(\R^d_y)$. Using this applied to a function $\varphi$, we
recover \eqref{eq:vector-direct-integral}. In the main text, the operator $\Pi$ acting on the full $\rL^2$ is the direct integral
of the operators $\Pi (x)$ that act on the electronic $\rL^2$ space, namely $\rL^2_y$.\\
The strange integral notation may be understood in the following situation.
Replace the above Hilbert space $\rL^2(\R^n_x)$ by some finite dimensional space $\C^p$. Then 
\begin{equation*}
\rL^2(\C^p_x; \rL^2(\R^d_y))\ =\ \int^{\oplus}\rL^2(\R^d_y)\, dx
\end{equation*}
is actually a finite direct sum of copies of the space $\rL^2(\R^d_y)$, the elements of which are just $p$-uple $(\psi _1, \cdots , \psi _p)$ of $\rL^2(\R^n_y)$-functions. By passing from this finite dimensional case to the infinite dimensional one, it is natural to replace the finite direct sum by a ``direct integral''. 

{\em True and generalized eigenfunctions.} The operator $H$ in the text is considered as a self-adjoint operator in $\rL^2$ (in all variables).
A true eigenfunction is a $\rL^2$, nonzero function $\varphi$ in $D(H)$ such that, for some
$\lambda$, $H\varphi=\lambda \varphi$. Since $H$ is also a differential operator, one may have
(many) solutions of the previous equation that do not belong to $\rL^2$. Such a solution is called
a generalized eigenfunction of $H$. It may have a quite bad regularity. 

{\em Distributions.} A distribution $T$ on $\R^d$ is a continuous linear form on the smooth, complex valued functions with compact support.
This means that, for such functions $f$ and $g$, for complex numbers $a$ and $b$, the function $af+bg$ is still smooth and has compact support and
the complex number $T(af+bg)$ is equal to $aT(f)+bT(g)$. The continuity property is a bit involved. 
To a (locally) integrable function $\varphi$ on $\R^d$, one can associate
a distribution $T$ defined by:
\[f\, \donne\, T(f)\ =\ \int_{\R^d}\varphi (x)f(x)\, dx\period\]
However there are distributions that are not associated to any function. This is the case of the Dirac distribution at
$x_0\in\R^d$, which is defined by $f\donne f(x_0)$. One can define many operations on distributions like the sum, the
multiplication by a smooth function, the differentiation, the Fourier transform (on so-called tempered distributions) but
the convolution and especially the product are only partially defined. Here we mean that one has to choose appropriately
the two distributions entering in the product (or the convolution). A theory of distributions can be found in \cite{t1}.

{\em Pseudodifferential operators.} A pseudodifferential operator on $\rL^2(\R^d)$ is an operator formally given by
$f\donne Af$ where $Af$ is a new function on $\R^d$ defined by 
\[Af (x) = (2\pi)^{-d}\int e^{i(x-y)\cdot\xi}a(x, \xi)f(y)\, dyd\xi\comma\]
for some appropriate function $a(x, \xi)$. If $a$ identically equals one, one see, using the
Fourier and the inverse Fourier transforms, that $A$ is the identity operator $f\donne f$.
If $a(x, \xi )$ equals the norm of $\xi$ squared then one can check in the same way that $A$ is
the Laplace operator $\Delta_x$. Pseudodifferential operators are a generalization of
differential operators. If the function $a$ is matrix-valued (and $f$ vector-valued) then $A$
is a pseudodifferential matrix operator. An introduction to pseudodifferential operators is
provided in \cite{t2}.


\begin{tikzpicture}[scale=1.2]
%
\draw[->] (0,0)--(10,0);
\draw[->] (0,-3)--(0,10);
\path(0.2,10) coordinate(src);
\path (10,3) coordinate(dst);
\path(2,-10) coordinate(ctrl1);
\path(1,2) coordinate(ctrl2);
\draw(src) .. controls(ctrl1) and (ctrl2) .. (dst);
\path(0.4,10) coordinate(src);
\path (10,4) coordinate(dst);
\path(2,-5) coordinate(ctrl1);
\path(5,6) coordinate(ctrl2);
\draw(src) .. controls(ctrl1) and (ctrl2) .. (dst);
\path(0.6,10) coordinate(src);
\path (10,4.3) coordinate(dst);
\path(3,-4) coordinate(ctrl1);
\path(7,6) coordinate(ctrl2);
\draw(src) .. controls(ctrl1) and (ctrl2) .. (dst);
\path(1,10) coordinate(src);
\path (10,7) coordinate(dst);
\path(4,5) coordinate(ctrl1);
\draw[top color=blue](src) .. controls(ctrl1)  .. (dst)--(10,10)--cycle;
\path (2.1,+.3) node {$|x_0|$};
\fill[color=black](2.1, -1.49)circle(.5mm);
\draw[dashed] (2.1,0)--(2.1, -1.5);
\draw[->] (3.1, -1.95) --(2.2, -1.55);
\path (3.6, -1.95) node {$\inf \lambda_1$};
\draw[dashed](0, -1.6)--(5, -1.6);
\path(-.3, -1.2) node {$E^0_+$};
\draw[dashed](0, -1.4)--(5, -1.4);
\path(-.3, -1.8) node {$E^0_-$};
\draw[dashed](6,0)--(6,5.8);
\draw[thick] (6,5.8)-- (6,10);
\path(4.6, 8) node {$\sigma_c(Q(x^*))$};
\draw[->](5.4, 8)-- (5.9, 8);
\path(6, -0.3) node{$|x^*|$};
\fill[color=black](6, 1.6)circle(.5mm);
\draw[->](5, 1.6)-- (5.9, 1.6);
\path(4.5, 1.6) node{$\lambda_1(x^*)$};
\fill[color=black](6, 2.95)circle(.5mm);
\draw[->] (5, 2.95) --(5.9, 2.95);
\path(4.5, 2.95) node{$\lambda_2(x^*)$};
\fill[color=black](6, 3.33)circle(.5mm);
\draw[->] (5, 3.6) --(5.9, 3.4);
\path(4.5, 3.6) node{$\lambda_3(x^*)$};
\draw[dashed](0, 2.2)--(10, 2.2);
\path(-.3, 2.4) node {$E_+$};
\draw[dashed](0, 2)--(10, 2);
\path(-.3, 1.8) node {$E_-$};
\draw[dashed](6,0)--(6,5.8);
\draw[dashed](7.4,0)--(7.4, 6.15);
\draw[thick] (7.4, 6.15) -- (7.4, 10);
\path(8.8, 8) node {$\sigma_c(Q(x_*))$};
\draw[<-](7.5, 8)-- (7.95, 8);
\path (7.4,-.3) node {$|x_*|$};
\fill[color=black](7.4, 2.34)circle(.5mm);
\draw[<-](7.5, 2.3)-- (8, 1.8);
\path (8.4, 1.6) node {$\lambda_1(x_*)$};
\fill[color=black](7.4, 3.82)circle(.5mm);
\draw[<-](7.5, 3.7)-- (8, 3.2);
\path (8.4, 3) node {$\lambda_2(x_*)$};
\fill[color=black](7.4, 4.01)circle(.5mm);
\draw[<-](7.5, 4.1)-- (8, 4.8);
\path (8.4, 4.9) node {$\lambda_3(x_*)$};
\fill[color=black](8.06, 4.18)circle(.5mm);
\draw[<-](8.1, 4.1)-- (9.2, 3.6);
\path (10.7, 3.5) node {eigenvalue crossing};
\path (10 ,-.3) node {$|x|$};
\path (6,-5) node {\bf Figure $1$: spectra of electronic Hamiltonians};
\end{tikzpicture}

\newpage


\begin{tikzpicture}[scale=1.2]
%
\draw[->] (0,0)--(10,0);
\draw[->] (0,-3)--(0,10);
\path(0.2,10) coordinate(src);
\path (10,0.5) coordinate(dst);
\path(2,-11) coordinate(ctrl1);
\path(1,1) coordinate(ctrl2);
\draw(src) .. controls(ctrl1) and (ctrl2) .. (dst);
\path(0.4,10) coordinate(src);
\path (10, 2.7) coordinate(dst);
\path(2,-8.5) coordinate(ctrl1);
\path(3,4) coordinate(ctrl2);
\draw(src) .. controls(ctrl1) and (ctrl2) .. (dst);
\path(0.6,10) coordinate(src);
\path (10,3.1) coordinate(dst);
\path(3,-6.6) coordinate(ctrl1);
\path(8,6) coordinate(ctrl2);
\draw(src) .. controls(ctrl1) and (ctrl2) .. (dst);
\path(0.8,10) coordinate(src);
\path (10,5) coordinate(dst);
\path(3,5) coordinate(ctrl1);
\path(5,6) coordinate(ctrl2);
\draw(src) .. controls(ctrl1) and (ctrl2)  .. (dst);
\path(1,10) coordinate(src);
\path (10,5) coordinate(dst);
\path(2,2) coordinate(ctrl1);
\path(5,8) coordinate(ctrl2);
\draw(src) .. controls(ctrl1) and (ctrl2) .. (dst);
\path(1.2,10) coordinate(src);
\path (10,7) coordinate(dst);
\path(4,6) coordinate(ctrl1);
\draw[top color=blue](src) .. controls(ctrl1)  .. (dst)--(10,10)--cycle;
\draw[dashed] (10,0)--(10, 7);
\fill[color=black](10, 2.5)circle(.5mm);
\draw[<-](10.05, 2.48)-- (11, 2);
\path (11.2, 1.9) node {$E^c$};
\fill[color=black](10, 2.7)circle(.5mm);
\draw[<-](10.05, 2.7)-- (10.75, 2.7);
\path (11.2, 2.7) node {$E^c(0)$};
\fill[color=black](10, 4)circle(.5mm);
\draw[dashed] (0,4)--(10, 4);
\draw[<-](10.05, 4.01)-- (10.95, 3.7);
\path (11.2, 3.6) node {$E^c_-$};
\fill[color=black](10, 4.4)circle(.5mm);
\draw[dashed] (0,4.4)--(10, 4.4);
\draw[<-](10.05, 4.41)-- (10.95, 4.85);
\path (11.2, 4.8) node {$E^c_+$};
\draw[<-](10, -0.1)-- (10.9, -0.8);
\path (11.5 ,-1) node {infinity};
\path (9.5 ,-.3) node {$|x|$};
\fill[color=black](5.02, 1.27)circle(.5mm);
\draw[<-](5.07, 1.24)-- (7.2, -2.2);
\fill[color=black](7.2, 2.5)circle(.5mm);
\draw[<-](7.21, 2.45)-- (7.5, -2.2);
\path (7.7, -2.5) node {eigenvalue crossings};
\path (6,-5) node {\bf Figure $2$: scattering situation};
\end{tikzpicture}

\newpage


%
%
\end{document}